\documentclass[journal]{IEEEtran}

\usepackage{graphicx}
\graphicspath{{figs/}}
\usepackage{times} 
\usepackage{amsmath} 

\usepackage{xcolor} 
\usepackage{url}
\usepackage{amssymb}
\usepackage{mathtools}
\usepackage{breqn}
\usepackage[ruled,vlined,linesnumbered,lined,boxed,commentsnumbered]{algorithm2e}

\begin{document}

 \pagebreak
 
This work has been submitted to the IEEE for possible publication. Copyright may be transferred without notice, after which this version may no longer be accessible.
 
  \pagebreak

\title{Motion Estimation for Optical Coherence Elastography Using Signal Phase and Intensity}

\author{Hossein~Khodadadi, Orcun~Goksel, Sabine~Kling
\thanks{Department of Information Technology and Electrical Engineering, ETH~Zurich, Switzerland.}

}

\maketitle

\begin{abstract}
Displacement estimation in optical coherence tomography (OCT) imaging is relevant for several potential applications, e.g.\ for optical coherence elastography (OCE) for corneal biomechanical characterization.
Larger displacements may be resolved using correlation-based block matching techniques, which however are prone to signal de-correlation and imprecise at commonly desired sub-pixel resolutions. 
Phase-based tracking methods can estimate tiny sub-wavelength motion, but are not suitable for motion magnitudes larger than half the wavelength due to phase wrapping and the difficulty of any unwrapping due to noise.
In this paper a robust OCT displacement estimation method is introduced by formulating tracking as an optimization problem that jointly penalizes intensity disparity, phase difference, and motion discontinuity.
This is then solved using dynamic programming, utilizing both sub-wavelength-scale phase and pixel-scale intensity information from OCT imaging, while inherently seeking for the number of phase wraps.
This allows for effectively tracking axial and lateral displacements, respectively, with sub-wavelength and pixel scale resolution.
Results with tissue mimicking phantoms show that our proposed approach substantially outperforms conventional methods in terms of axial tracking precision, in particular for displacements exceeding half the imaging wavelength.
\end{abstract}


\IEEEpeerreviewmaketitle
\newcommand{\eg}{\textit{e.g.}, }
\newcommand{\ie}{\textit{i.e.}, }

\section{Introduction}
Optical Coherence Elastography (OCE)~\cite{kennedy2017emergence,Larin2017} is of interest for estimating tissue strain and monitoring dynamic deformation responses in both, \textit{ex-vivo} biological samples \cite{wang2014noncontact} and in \textit{in-vivo} corneal tissue \cite{manapuram2012vivo, de2018live}. 
Current methods for estimating deformations in OCE are based either (i)~on speckle tracking, \eg relying on cross-correlation \cite{zaitsev2013correlation,zaitsev2015,de2018live} which is prone to speckle \textit{boiling} and \textit{blinking}~\cite{zaitsev2015}, or (ii)~on phase difference estimation \cite{manapuram2012vivo, wang2014noncontact}, which typically suffers from phase wrapping when the displacements exceed half the wavelength. 

Correlation-based speckle tracking methods typically resolve pixel-level displacements, but they are inaccurate at estimating smaller motions. 
These methods also suffer from speckle decorrelation, which is significant in OCT imaging due both to the interferometric (coherence) nature of OCT speckles leading to strain-induced speckle boiling/blinking, and to the relatively smaller OCT speckle sizes (down to 1-2 pixels) compared to other imaging such as ultrasound. 
Consequently, speckle-tracking approaches have practically not been employed in OCE, despite over 15 years of effort in this direction~\cite{zaitsev2015}.
In contrast, phase-based methods are ideally suited for strain computation in OCE due to their better tolerance to strain-induced speckle blinking and boiling, and intrinsically higher sensitivity to (sub-wavelength) displacements which allows measuring smaller displacements unambiguously. 
Note that the displacements that can be measured from the phase directly, \ie without any phase unwrapping, is limited to a quarter of the wavelength, \ie $\lambda_0/4 \approx 0.2\,\mu$m for the typical central wavelength of an OCT source. 
Assuming a typical OCT image depth of 2\,mm, the above limit means that only the strains smaller than $\approx 0.1$\% can be unambiguously observed using na\"ive phase-based methods.
To overcome such limit, phase unwrapping can be applied in a cumulative fashion at incremental depths or displacements.
However, any inherent OCT measurement noise often easily corrupts the phase readings, leading to errors that ``break'' the phase at $\pm\pi$\,rad, detrimental to the unwrapping procedure and hence the estimated displacements.

In OCE, strain-produced displacements within a given OCT A- or B-scan may vary largely, from sub-pixel sub-wavelength values to those exceeding not only the wavelength, but also the pixel scales (that in itself may wrap several phases). 
Conventional phase-based OCT displacement estimation methods compare the phases of reference and deformed (pre- and post-deformation) scans at a given pixel location~\cite{wang2007phase,kennedy2012strain,chin2014analysis}. 
For very small displacements, it is reasonable to assume such pixel to contain mostly similar scatterers in reference and deformed scans, so that the phase variations can be directly related to scatterer displacements. 
However, with supra-pixel displacements the same pixel location of the two scans may contain arbitrarily different scatterers, thus the phases of the two OCT signals would be completely unrelated, with their difference essentially yielding a random value.
Moreover, although small deformations may help avoid phase wrapping, 
these in turn lead to motion closer to the noise floor and thus a lower signal-to-noise ratio in any subsequent strain estimation. 
For the above reasons and practical concerns where small deformations cannot be ensured, being able to estimate larger displacements and hence strains is very desirable.  
To that end, adjustment schemes for consistent phase retrieval~\cite{zaitsev2016optimized,matveyev2018vector,zaitsev2016hybrid,zaitsev2016robust} have been introduced. 
In \cite{zaitsev2016robust,zaitsev2016hybrid}, local displacements are accumulated to correct for supra-pixel displacements when calculating phase-variations.
Local strains are determined by the local phase gradient over a chosen vertical processing window. 
Although this approach circumvents phase wrapping when for moderate local strains ($\approx$ 5\%), a direct strain computation is still prone to integration of errors in the displacement estimation, particularly problematic with supra-pixel displacements. 
In~\cite{zaitsev2016optimized,matveyev2018vector}, the above idea is extended with the phase retrieved by averaging the complex-valued OCT signal, as opposed to the earlier methods which use only the phase information. 
This extension, however, only improves the phase-variation estimation but it does not solve the error integration problem. 
None of the above phase-based methods neither estimates nor takes into account lateral displacements, which is a major limitation where any tissue location may move laterally. 
Moreover, axial translation is often assumed to be zero, 
although (since the OCT probe is non-contact, unlike, \eg in ultrasound elastography) axial translations may occur and if not corrected for, may degrade the displacement estimations substantially.

Optimization-based methods have been utilized in ultrasound elastography~\cite{j2011recent,Rivaz11_TMI,yuan2015analytical,ISBI2018} and these were shown to typically outperform cross-correlation methods.
In this work, we extend such techniques to OCT, leading to a novel method in OCT motion estimation by combining the advantages of correlation-based and phase-based displacement estimation approaches. 
We introduce, for the first time to our knowledge, an optimization-based method for displacement tracking in OCT images, by including motion continuity priors and therefore being more robust to signal decorrelation compared to previous approaches. 
The proposed method exploits both amplitude and phase information of complex valued OCT B-scans, to estimate axial and lateral displacement fields. 
Specifically, we apply a modified version of the vector-based phase gradient estimation method~\cite{matveyev2018vector} in a dynamic programming (DP) optimization scheme.
By incorporating phase wrapping in the cost function, our DP algorithm is designed to find the number of phase wraps as well as the measured phase difference between reference and deformed OCT scans by minimizing motion discontinuity and intensity disparity. 

\subsection{Modeling the OCT signal}
\label{sec:model}

Below we first present the OCT imaging model in~\cite{zaitsev2014model} and then generalize this to introduce our proposed displacement estimation algorithm.
Consider a single OCT A-line in the direction of z-axis, aligned with the probing-beam axis. 
In~\cite{zaitsev2014model}, the probing beam was considered to be a weakly diverging beam and to have uniformly distributed amplitude over a constant radius cylinder form, \ie without strong focusing. Consequently, the phase of the received-wave is entirely determined by the phase delay accumulated during the propagation forth-and-back in the axial direction. This means that for each spectral harmonic with the wavenumber $k_n$ where $n=1,2,...,N$, the received complex amplitude $B(z_s,k_n)$ is proportional to the incident field and scattering strength: 
\begin{equation}
\label{eq:OCTmodel}
    B(z_s,k_n)=R_s(k_n) I(k_n) e^{i2k_n z_s},
\end{equation}
where $z_s$ is the axial coordinate of scatterer $s$\,, $I(k_n)$ is the incident field amplitude, and $R_s(k_n)$ describes the scattering strength of $s$. 
The complex-valued amplitude of pixel $q$ in an A-line having depth $H$ and consisting of $N$~pixels enumerated as \mbox{$q$$=$$1,...,N$} would be:
\begin{equation}
\label{eq:OCTmodel1}
    A(z_q)=\sum_n\sum_s B(z_s,k_n)e^{-i\frac{2\pi n}{H}z_q},
\end{equation}
where $z_q=\frac{qH}{N-1}$ for $1$$=$$1,...,N$$-$$1$ is the coordinate of pixel $q$ within the A-line.
The light source is typically assumed to have a Gaussian-shaped spectrum with standard deviation $\sigma_k$$=$$\frac{\Delta k}{2\sqrt{2\ln{2}}}$ centered around a center wavenumber ${{k}_{0}}$. 
Accordingly, the incident field amplitude can be represented as:
\begin{equation}
    \label{eq:spectrum}
    I(k_n)=\frac{1}{\sqrt{2\pi \sigma_k^2}}\exp(-\frac{(k_n-k_0)^2}{2\sigma_k^2}).
\end{equation}
where ${{k}_{n}}$ are the spectral lines in the probing spectrum.
Since the interval $\Delta k ={{k}_{n+1}}-{{k}_{n}}$ between neighbouring spectral lines and the maximal depth $H$ in the spectral-domain OCT are related as $\Delta k= \pi/H$, then ${{k}_{n}}={{k}_{0}}+n\pi /H$ where $n=-\frac{N-1}{2},...,\frac{N-1}{2}$ assuming an odd number $N$ for symmetry. 
By replacing this in Eq.~\eqref{eq:OCTmodel} and by taking into account the dependence of the illuminated-beam amplitude on the lateral coordinate, Eq.~\ref{eq:OCTmodel1} can be rewritten as
\begin{equation}
\label{eq:OCTmodel2}
    A(z_q)=\sum_n\sum_s R_s(k_n) I(k_n)e^{i2k_0z_s}e^{i\frac{2\pi n}{H} z_s}e^{-i\frac{2\pi n}{H}z_q}.
\end{equation}
Herein the dispersion of $R_s(k_n)$ on different wavenumbers $k_n$ is assumed to be negligible. 

The amplitude of received OCT signal is affected both by absorption during the forth-and-back propagation and by divergence of the backscattered signals. 
This can be modeled in scatterer strength (reflectivity) as:
\begin{equation}
\label{eq:atten}
    R_s=p_0 e^{-\mu_t z_s},
\end{equation} 
with the light power $p_0$ of the source and the total attenuation coefficient $\mu_t=\mu_a+\mu_s$ consisting of absorption $\mu_a$ and scattering $\mu_s$ attenuation coefficients of the sample under inspection~\cite{bashkatov2005optical}. 

One of the limiting factors in spectral-domain OCT is the discretization and digitization in the detection process. 
These processes restrict the resolution and the probing depth of the system.
Spectral-domain OCT systems utilize a spectrometer that integrates the spectrum over a square pixel of the camera with width and separation $\Delta k$. 
Such integration leads to a convolution in $k$-domain with the filter
\begin{equation}  \label{eq:kfilterPix}
    G(k)=\frac{1}{\Delta k}\ \textsf{rect}\!\left(\frac{k}{\Delta k}\right)\,,
\end{equation}
with \textsf{rect} being the rectangle function~\cite{goodman2005introduction}. 
In addition, the spectrometer operates with a spectral resolution, typically described by a Gaussian with standard deviation $\sigma_r$. 
Such resolution integration leads to a convolution in $k$-domain with the filter
\begin{equation}  \label{eq:kfilterRes}
    F(k)=\frac{1}{\sqrt{2\pi \sigma_r^2}}\exp\!\left(-\frac{k^2}{2\sigma_r^2}\right)\,.
\end{equation}
These two filters in $k$-domain equates to a multiplication of the $z$-domain OCT signal by the functions $g(z)=\frac{1}{2\pi}\text{sinc}(\frac{\Delta k z}{\pi})$ and $f(z)=\frac{1}{\sqrt{2\pi}}\exp(-2z^2\sigma_r^2)$\,, respectively.

\subsection{Effect of scatterer motion on OCT signal}
\label{seq:scat_motion}
\begin{figure}
	\centering
		\includegraphics[width=.8\linewidth]{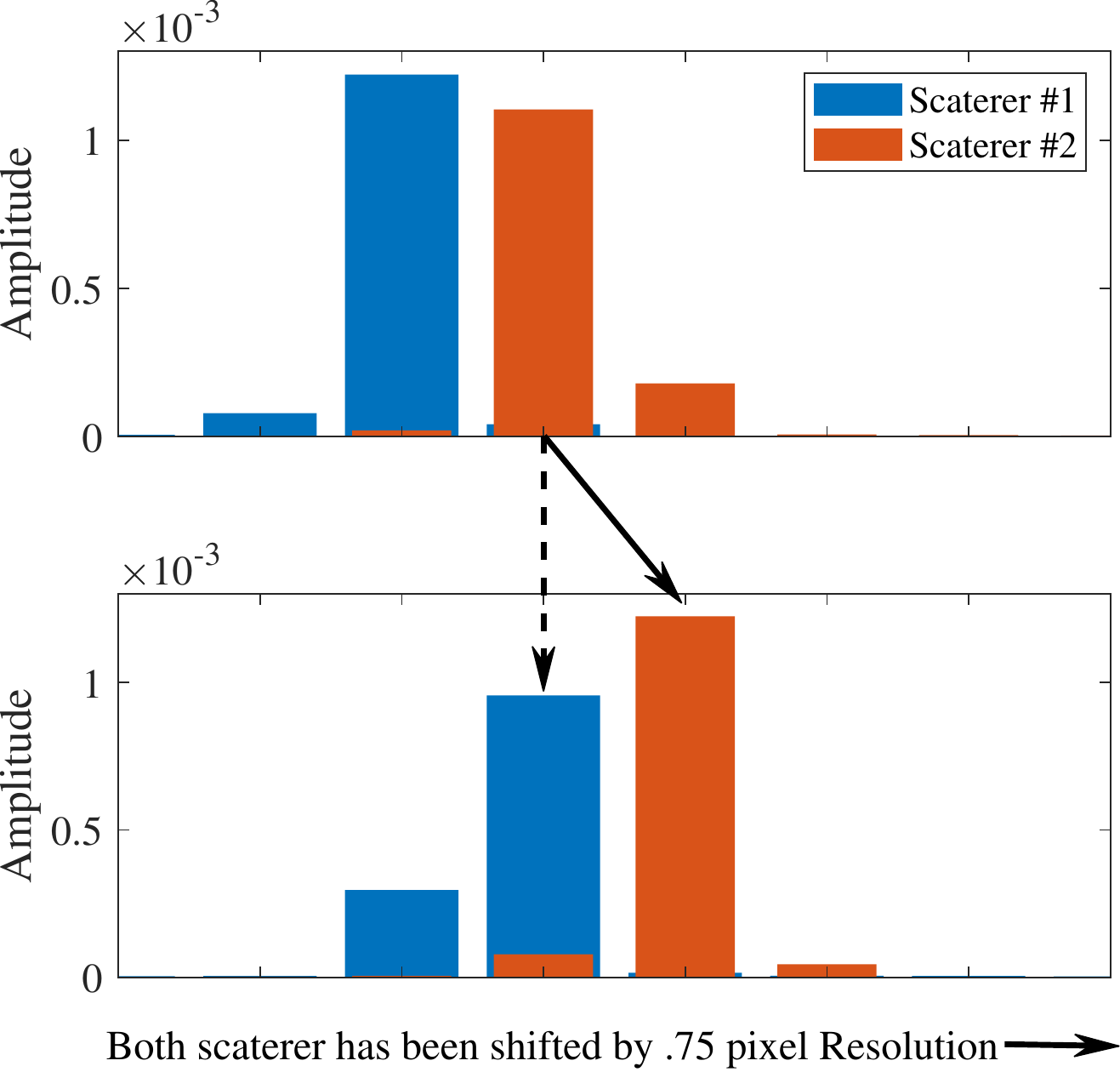}\\(a)\\[1ex]
		\includegraphics[width=.8\linewidth]{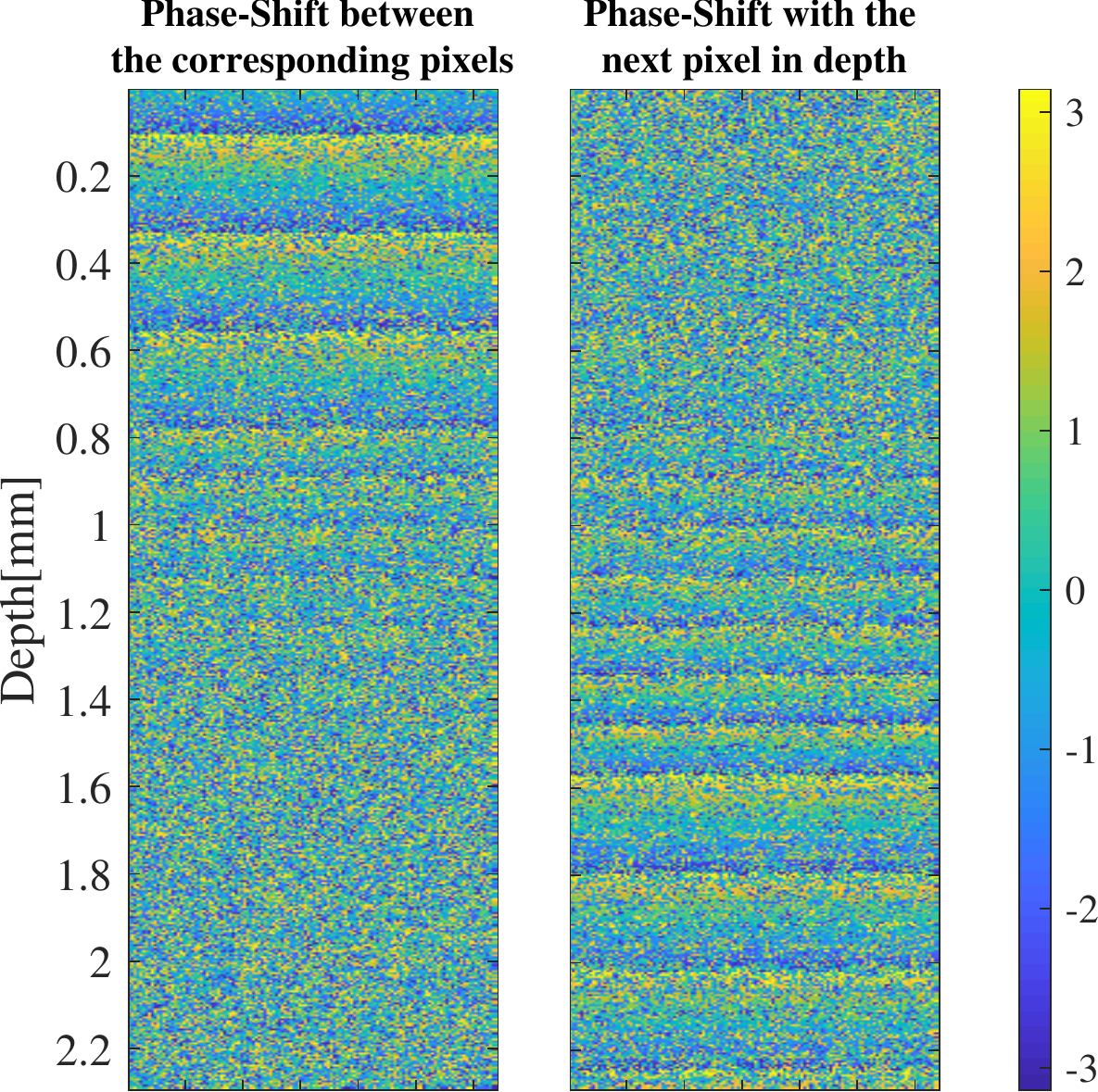}\\(b)
	\caption{Effect of scatterer motion on the OCT signal. Amplitude of a simulated OCT signal containing two scatterers with the same strength and $1.25$ pixel resolution distance from each other is presented in the top graph of (a). These two scatterers are then shifted down by $0.75$ pixel resolution in the depth (right direction in the plot). (b)-left shows the phase-shift between corresponding pixels between a simulated original and deformed OCT signal, while (b)-right shows the phase-shift of each pixel in the original signal and one pixel deeper in depth in the deformed signal. 
	}
	\label{fig:sim_theory}
\end{figure}

In contrast to the ideal rectangular spectral shape of the source, which can ensure an ultimately narrow localization within a single pixel number, a smoother spectrum Eq.~\eqref{eq:spectrum} usually introduces a spatial smoothing of imaged point-like particles.
Thus, speckles are spread between at least two adjacent pixels. 
This causes complications in any phase-based displacement estimation: 
For instance, consider the case illustrated in Fig.\,\ref{fig:sim_theory}(a) showing two scatterers with the same strength and $1.25$ pixels ($=$$5.6 \mu$m) away from each other.
If the tissue displaces by simply translating both these pixels in depth (right direction in the plot), \eg by $0.75$ pixel ($3.36\mu$m), calculating the phase-shift using the same pixel (the dashed arrow) would produce a false estimation, with the phase of an unrelated (neighbouring) scatterer now mostly occupying this pixel (and their partial interference further corrupting each other). 
This effect is also demonstrated with simulated OCT signals in Fig.\,\ref{fig:sim_theory}(b)-left, which shows the phase-shift in each pixel between a reference and deformed OCT frames.
The scatterers in the reference were compressed axially by one pixel from the bottom side upwards before simulating the deformed frame; using the two frames the pixel-wise phase-shift displayed in the figure.
It is seen that as the displacement magnitudes increase downwards and reaches supra-pixel values, a na\"ive phase-based approach fails to track displacements and given such errors, any attempt to use these to compensate for accumulated displacements (and hence pixel shifts) would also fail in turn. 
Fig.\,\ref{fig:sim_theory}(b)-right shows the phase-shift from each pixel in the reference signal to one pixel deeper in the deformed signal, which indicates the tracking potentially recoverable in depth, if such cumulative pixel shift can be anticipated.
Hence, methods that compensate for accumulated displacements in depth can use all phase variation information (both in the left and right sub-figures in this example) when estimating displacements, \eg in the respective halves of the shown simulation.  

\section{Methods}

Consider two consecutive complex valued OCT B-scans $I_1[i,j]=\beta_1[i,j]e^{\imath\phi_1[i,j]}:I_1 \in \mathbb{C}^{m\times n}$ and $I_2[i,j]=\beta_2[i,j]e^{i\phi_2[i,j]}:I_2 \in \mathbb{C}^{m\times n}$  collected before and after the sample undergoes deformation. Similarly to the medical image registration literature, an energy based solution for the deformable image alignment can be formulated as an optimization problem with the following cost function:
\begin{equation}
	\label{eq:c1}
	\min_{A,L}P=\min_{A,L}[\Delta(I_1,I_2,A,L)+\alpha\Psi(A,L)] 
\end{equation}
where the first term $\Delta(I_1,I_2,A,L)$ is a penalty function for speckle (\ie intensity) decorrelation,  the second term $\Psi(A,L)$ is a penalty function for losing motion continuity, and $\alpha$ is the regularization weight. Based on this optimization, the two matrices ($A$ and $L$) describing axial and lateral displacements, respectively, shall be determined such that $I_1[i,j]\sim I_2(i+a_{i,j},j+l_{i,j})$ where $a_{i,j}$ and $l_{i,j}$ are the entries of matrices $A$ and $L$, respectively. Note that calculating $I_2(i+a_{i,j},j+l_{i,j})$ requires a sub-pixel interpolation of $I_2$ if $a_{i,j}$ and $l_{i,j}$ are not integers. 
This optimization problem can be solved by the Viterbi algorithm, which is a dynamic programming technique based on Bellman's principle of optimality~\cite{bellman1954}. 
This algorithm allows for solving complex problems by breaking them into a collection of simpler sequential sub-problems, where each of them corresponds to a discrete decision. 

While both intensity and phase information are available in the axial direction in OCT, in the lateral direction only the intensity information is available for displacement estimation. Here, the phase information arises from the phase difference $\Phi[i,j]=\phi_1[i,j]-\phi_2[i,j]$ between the reference and deformed OCT B-scans. 
The phase difference is deterministic and encodes axial displacements as follows:
\begin{equation}
	\label{eq:disp}
	U[i,j]=\frac{\lambda_0\Phi[i,j]}{4\pi r_n},
\end{equation}
where $\lambda_0$ is the center wavelength of the OCT beam and $r_n$ is the refractive index of the sample. Since the phase difference is between $[-\pi,\pi]$ and may wrapped around, the displacement information using the above alone is ambiguous as soon as it exceeds $\frac{\lambda_0}{4 r_n}$. Larger displacements can still be retrieved by applying an unwrapping algorithm; however, this is highly sensitive to noise.
In this regard, complex vector averaging~\cite{matveyev2018vector} in lateral direction may help suppress noise from small-amplitude pixels, but at the cost of reducing lateral resolution. 
We propose below to use DP to address such problem of phase unwrapping.

\subsection{Dynamic Programming Formulation}

DP is designed to find the global optimum of a cost in a discrete decision space, which in our formulation consists of speckle disparity and motion continuity. Let us define $A\in\mathbb{R}^{m\times n}$ and $L\in \mathbb{R}^{m\times n}$ as the discretized axial and lateral displacement matrices (the out-of-plane motion is not considered herein), where each of their elements $a_{i,j}$ and $l_{i,j}$ satisfy $\|a_{i,j}\|\le a_{\max}$ and $\|l_{i,j}\|\le l_{\max}$, and $a_{\max}$ and $l_{\max}$ are the maximum possible axial and lateral displacement estimates, respectively (bounded by the applied displacement amplitude). Assume that the set $\mathcal{A} \coloneqq \{a_r\mid  \|a_r \|< a_{\max}\}_{{r}=1}^{R}$ quantizes the axial displacement range into $R$ discrete values, and the set $\mathcal{L}\coloneqq \{l_{q} \mid  \|l_{q} \|< l_{\max}\}_{{q}=1}^{Q}$ partitions the lateral displacement range into $Q$ discrete values. We then denote the discrete decision space by
\begin{equation}
\mathcal{S}\coloneqq\{s_p:=(a_r,l_q)\mid  a_r \in \mathcal{A}, l_q \in \mathcal{L}\}_{{p}=1}^{R.Q}\ .
\end{equation}

According to Eq.~\eqref{eq:disp}, any potentially wrapped phase differences are related to axial displacements at each pixel. The discrete nature of DP decision space and the unknown number of phase wraps that may exist at each pixel motivated us for a solution to mitigate phase wrapping ambiguity by expressing the problem to relate the phase wraps to a decision space.
In particular, we herein propose to redefine the axial displacement set as
\begin{equation}
\mathcal{A}=\{a_r=\frac{r\lambda_0}{2r_n}: \|a_r \|< a_{\max}\}_{r=-\frac{R}{2}}^{\frac{R}{2}}\,,
\end{equation}
where each phase wrap is then represented by a state of the decision space in our DP. 
In other words, unlike the classical phase unwrapping methods which only use the information of the amplitude of jumps in phase, in this framework the number of phase wraps at each pixel is determined using the DP cost which includes the intensity information and motion regularization over the entire axial line. 
Note that with this, we can accommodate even phase jumps potentially more than $2\pi$. 
Moreover, the sought lateral displacement set $\mathcal{L}$ can also account for sub-pixel displacement values by lateral interpolation of intensity values. 
To further reduce phase noise, we adopted the vector-based method~\cite{matveyev2018vector} to laterally average the complex valued phase differences $\hat{\Phi}_{s_p}$ for each state $s_p$ of the DP as the normalized 2D cross-correlation as follows:
\begin{equation}
	\label{eq:phase}
	\beta[i,j]e^{\imath\hat{\Phi}_{s_p}[i,j]}=\sum\limits_{w=-W/2}^{W/2}I_1(i,j+w)I^*_2(i+\lfloor a_r\rfloor,j+\lfloor l_{q} \rfloor+w),
\end{equation}
where $W$ is an even number and $W+1$ determines the size of the lateral averaging window,  $\hat{\Phi}_{s_p}[i,j]$ is the laterally averaged phase difference, $I^*_2$ is the complex conjugate of $I_2$, and $\lfloor \cdot \rfloor$ denotes the floor function converting any subpixel displacements to pixel scale. 
For each DP state $s_p$, axial displacement is retrieved from phase difference according to Eq.~\eqref{eq:disp} 
as:
\begin{equation}
\label{eq:alpha_sp}
\alpha_{s_p}[i,j]=\frac{\lambda_0\hat{\Phi}_{s_p}[i,j]}{4\pi r_n}.    
\end{equation} 
Note that, even though the DP state $a_r$ is a multiple of $2\pi$, the displacement $\alpha_{s_p}$ may assume any continuous value given the phase difference.

To use DP to find the displacement matrices $A$ and $L$, we first define a intensity disparity term

\small
\begin{dmath}
	\label{eq:DP_disparity}
		D^{i,j}(d_a,d_l)= 
		\frac{\sum_{x}\sum_{y}(| I_1^{i+x,j+y}|-\mu_{1})(| I_2^{i+x+d_a,j+y+d_l}|-\mu_{2})}{\left [\sum_{x}\sum_{y}(| I_1^{i+x,j+y}|-\mu_{1})^2\sum_{x}\sum_{y}(| I_2^{i+x+d_a,j+y+d_l}|-\mu_{2})^2\right]^{\frac{1}{2}}}
\end{dmath}
\normalsize
where the sums over $x=[-w_1,...,w_1]$ and $y=[-w_2,...,w_2]$ cover a 2D cross-correlation processing window of the size $(2w_1+1)\times (2w_2+1)$ and $\mu_{i}$ is the mean value of scan $I_i$ within this window. 
Since axial and lateral displacements $d_a,d_l$ are not necessarily integers, to calculate Eq.~\eqref{eq:DP_disparity}, one needs to use 2D interpolation algorithms (bi-cubic interpolation is used in this work).

For spatial regularization, an axis-weighted finite-difference approach is used as follows:
\begin{equation}
	\label{eq:DP_Reg}
	R(d_a,d_l,d_a^-,d_l^-)=\beta \left \vert d_a - d_a^- \right \vert+\gamma\left \vert d_l - d_l^-\right \vert.
\end{equation}
where $d_a,d_l$ are axial and lateral displacements at pixel $[i,j]$ and $d_a^-,d_l^-$ are the axial and lateral displacements at the previous axial pixel $(i-1,j)$ respectively. 
Parameters $\gamma$ and $\beta$ are weights, respectively, for lateral and axial motion discontinuity penalties, which depend on the pixel resolution, the expected signal to noise ratio (SNR), and the expected maximum strain in each axis. 
Instead of the finite-difference regularization penalty above, one could also use 2nd-order differences, Laplacian, or other formulations, \eg to constrain strain continuity rather than displacement.

Combining Eqs.~\eqref{eq:DP_disparity} and \eqref{eq:DP_Reg}, the cost function at pixel $i$ of A-line $j$ can be defined as:
\begin{equation}
	\label{eq:cost}
		C^{i,j}_{s_p}= \left\vert D^{i, j}(\alpha_{s_p}^{i,j},l)\right \vert + \min _{s_p^-}  \left\{C^{i-1,j}_{s_p^-}+ R(\alpha_{s_p}^{i,j},l, \alpha_{s_p^-}^{i-1,j},l^-) 
		\right\},
\end{equation}
where $\alpha_{s_p}$ is calculated based to Eq.~\eqref{eq:alpha_sp} and the second term is a minimization over all DP states $s_p^-$. 
The cost function at each DP state $s_p$ is initialized for each A-line $j$ at $i=1$ with the intensity disparity term only. 
For $i=2,..., m$, the cost of preceding axial pixel is used to calculate the second term minimization, and the Viterbi algorithm is used to efficiently trace back the global optimum solution, \ie the accumulated minimum costs of reaching each state from its preceding pixel.
This procedure is elaborated in more detail in Algorithm~\ref{alg}.
\begin{algorithm}[t]\small
\SetAlgoLined
\KwData{$I_1,I_2,\mathcal{A},\mathcal{L},W,w_1,w_2,\beta, \gamma,\lambda_0,r_n.$}
\KwResult{Axial and lateral displacement fields $A$ and $L$.}
\For{each A-line $j\in \mathbb{N}_{\le n}$}{
\For{$s_p\in \mathcal{S}$}{
Calculate $\alpha_{s_p}^{1,j}$ according to Eqs.\ref{eq:phase} and \ref{eq:alpha_sp} for the first pixel $i=1$ in A-line $j$\;
Calculate $D^{1,j}(\alpha_{s_p}^{1,j},l)$  according to Eq.\ref{eq:DP_disparity}\;
Set $C^{1,j}_{s_p}= \left\vert D^{1, j}(\alpha_{s_p}^{1,j},l)\right \vert$ \;
}
\For{$i=2$ to $m$}{
\For{$s_p\in \mathcal{S}$}{
Compute $\alpha_{s_p}^{i,j}$ according to Eq.\ref{eq:alpha_sp},\;
\For{$s_p^{-}\in \mathcal{S}$}{
Calculate the spatial regularization using Eq. \ref{eq:DP_Reg},\;
Compute the transition cost $C^{i-1,j}_{s_p^-}+ R(\alpha_{s_p}^{i,j},l, \alpha_{s_p^-}^{i-1,j},l^-)$,
}
Find and store the minimizer $s_p^{-*}[i,j]$ of the transition cost (second term in Eq.\ref{eq:cost}) for each DP state $s_p$,\;
Calculate the intensity disparity term $D^{i, j}(\alpha_{s_p}^{i,j},l)$ using Eq. \ref{eq:DP_disparity},\;
Calculate $C^{i,j}_{s_p}$ according to Eq.\ref{eq:cost}, \;
}
}
Find the minimum of the cost $C^{m,j}_{s_p}$ at the last axial pixel $i=m$ over all DP states $s_p$. 
Output the entries $a_{m,j}=\alpha_{s_p^*}$ in A and $l_{m,j}=l_q^*$ in $L$ where the $s_p^*\coloneqq(a_r^*,l_q^*)$ is the minimizer.\;
Trace back all the minimizers using the stored values of $s_p^{-*}[i,j]$ and update the $A$ and $L$ entries accordingly.
} 
\caption{DP displacement tracking algorithm}
\label{alg}
\end{algorithm}

\section{Simulation Study}
To assess our proposed approach, we used the OCT imaging model described in Section~\ref{sec:model} to simulate OCT images and the strain-induced evolution of their speckle structure.  In our simulations we ignored any overlapping of A-lines in the lateral direction, so that scatterers in horizontally adjacent pixels are considered independent. The light source is assumed to have a power of $p_0=2.9$\,mW and the medium with a total  attenuation coefficient of $\mu_t=500$. 
The spectrum of the source is assumed to be Gaussian, with an assumed noise power of $p_n=2.9 \times10^{-13}$\,W (corresponding to a typical SNR of $100\,dB$). The average density of scatterers is set to be 2 per pixel. We assume that each A-line consists of $1024$ pixels, with the total depth of the image then being $H=2300\,\mu$m in air. 
The actual image depth would be half of this, \ie $512$ pixels, due to the mirroring effect inherent to spectral-domain OCT. The center wavelength of the optical source is chosen to be $878$\,nm with a spectral width of $62.5$\,nm, given the parameters of our experimental OCT hardware, which also are typical values for spectral-domain OCT scanners.

\begin{figure}
	\centering
		\includegraphics[width=.8\linewidth]{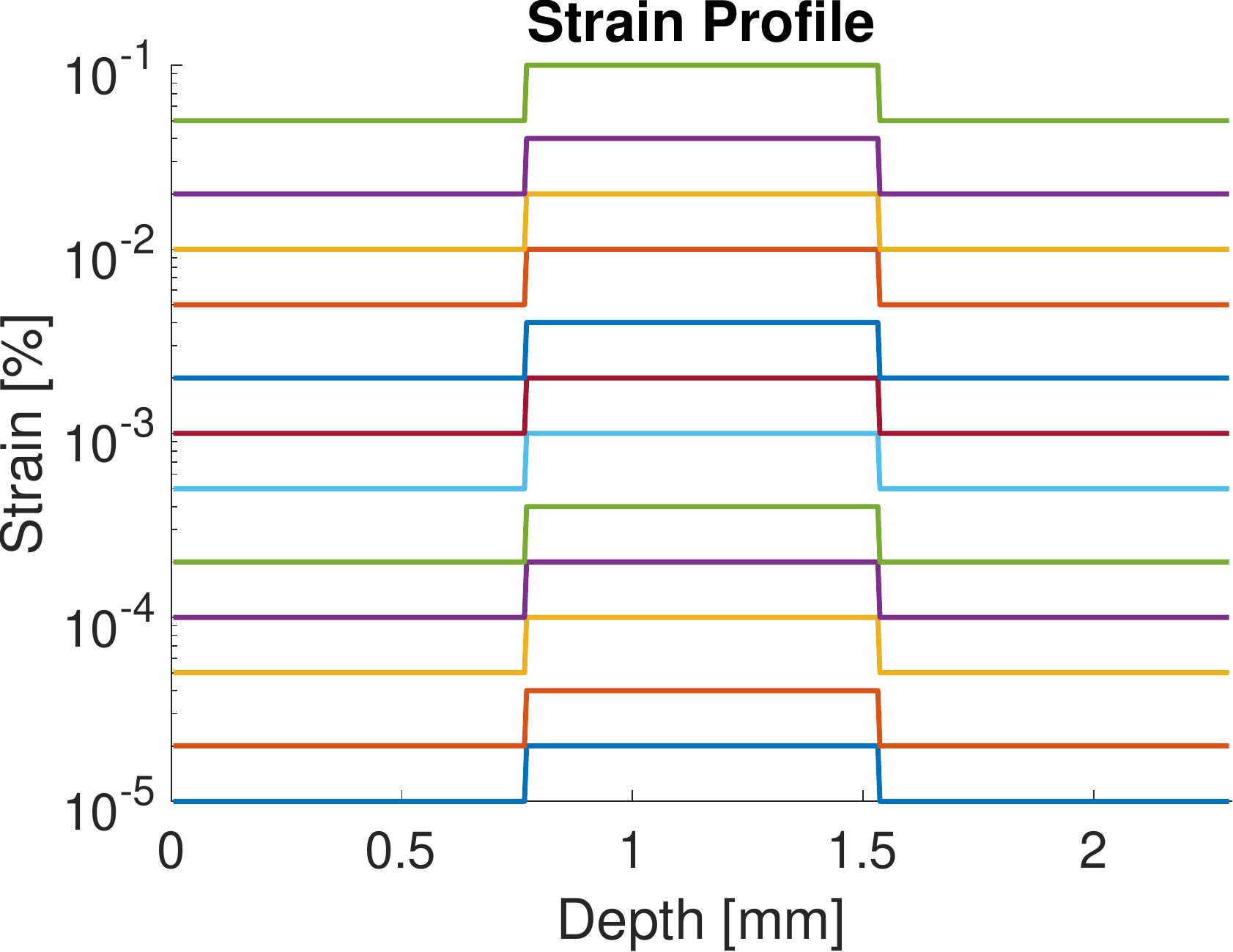}\\\qquad(a)\\[1ex]
		\includegraphics[width=.8\linewidth]{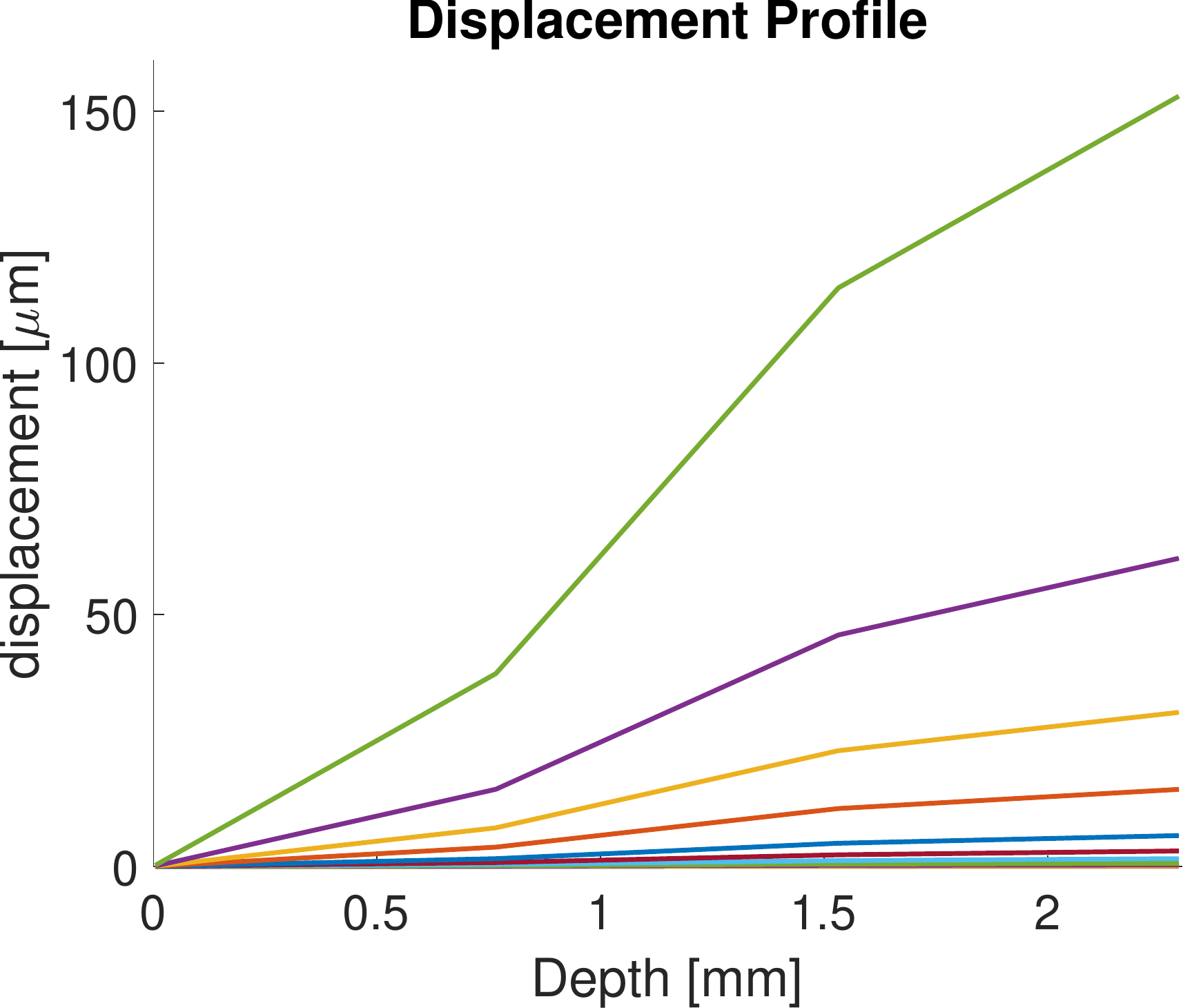}\\\qquad(b)
	\caption{Simulated strain (a) and displacement (b) profiles in depth. Same color strains and displacements indicate corresponding simulations.}.
	\label{fig:profile}
\end{figure}

To introduce OCE contrast in our simulated phantom, we set the material elasticity of axially the middle one-third of the simulated phantom, \ie $170 \le z \leq 340$ px, to be half the elasticity of the rest, such that for pure compressions the strain in this middle layer would be twice the rest. 
We first simulated one reference OCT B-scan with 128 A-lines.
Then, for different magnitudes of compressions each, we analytically computed the displacement at any reference scatterer location and then displaced all scatterers accordingly, before simulating a post-deformation OCT B-scan.
We repeated this separately from the reference, for 12 different amplitude values ($2\times10^{-5},4.31\times10^{-5},9.28\times10^{-5},2\times10^{-4},4.31\times10^{-4},9.28\times10^{-4},2\times10^{-3},4.31\times10^{-3},9.28\times10^{-3},2\times10^{-2},4.31\times10^{-2},9.28\times10^{-2}$), with which we aim to cover displacements from sub- to supra-pixels with a large strain range from $0.001\%$ to $10\%$.
Resulting analytically-computed strain and displacement profiles are shown in Fig. \ref{fig:profile}.
Note that for the maximum simulated strain, the largest displacements go up to 33 pixels ($150\,\mu$m).

To comparatively evaluate our proposed DP method, we compared it with those from conventional and state-of-the-art methods from different categories of displacement estimation techniques:
As a typical phase-based method, we used Kasai phase estimator~\cite{Kasai}.
As a typical intensity-based method, we used a cross-correlation (CC) speckle tracking algorithm.
As the state-of-the-art in phase-sensitive displacement/strain estimation, we used the vector approach to phase variation averaging (VP) in~\cite{matveyev2018vector}.
Representing a combination of intensity-based and phase-sensitive algorithms, we also devised a baseline (CC+VP) where large displacements are evaluated intensity-based by CC and then fine-tuned phase-based by VP.
Using each above method we computed the displacements and strains from the reference to each of the deformed OCT frames, and then quantitatively compared the results to known analytical form as the ground-truth.
Our method parameters were set to $a^{\text{max}}=150\mu m$, $W=20$, $\beta=\gamma=10^{-5}$, $w_1=w_2=5$. 

\begin{figure*}[h]
	\centering
	\begin{tabular}{cc}
		\includegraphics[width=0.48\textwidth]{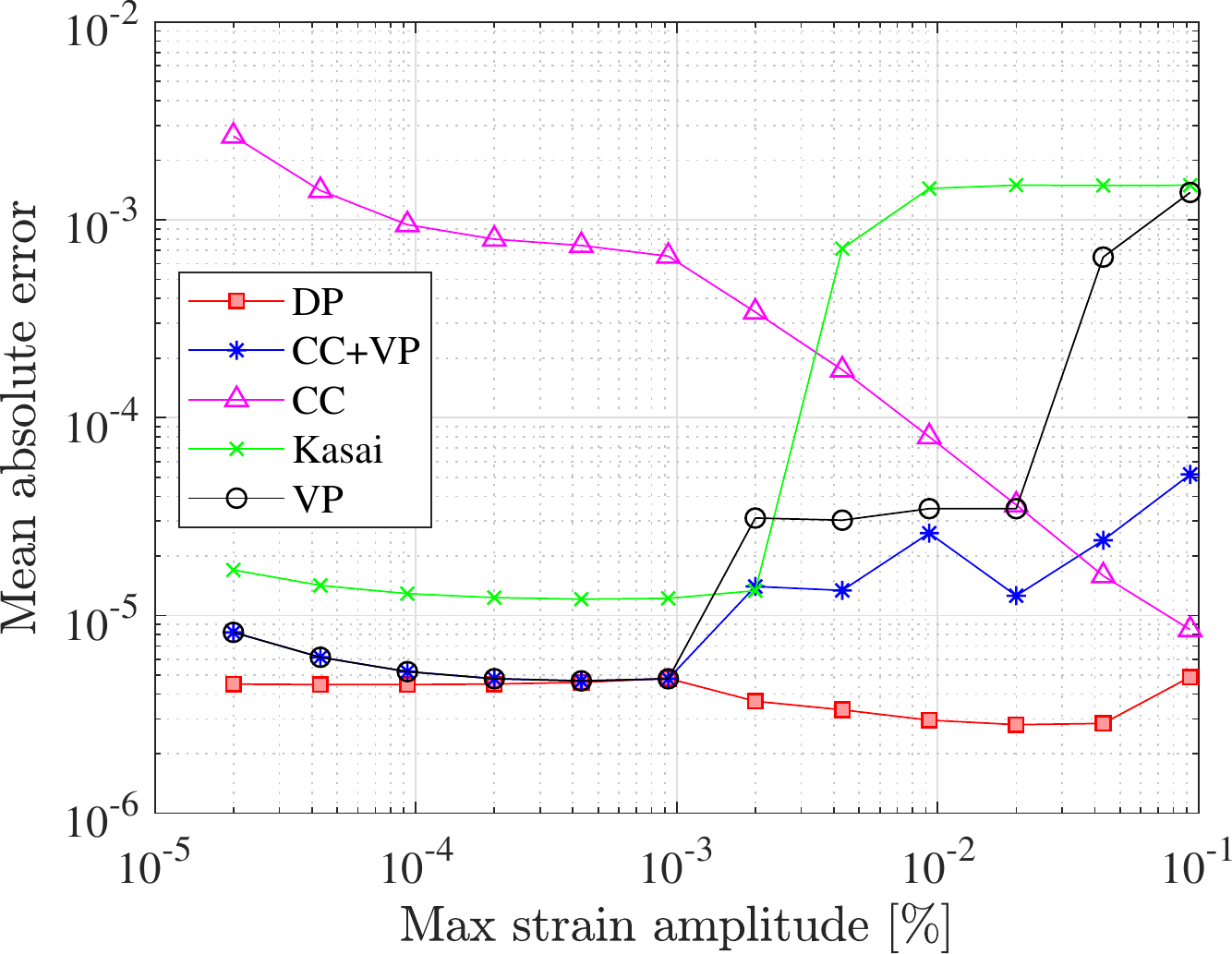}&\includegraphics[width=0.48\textwidth]{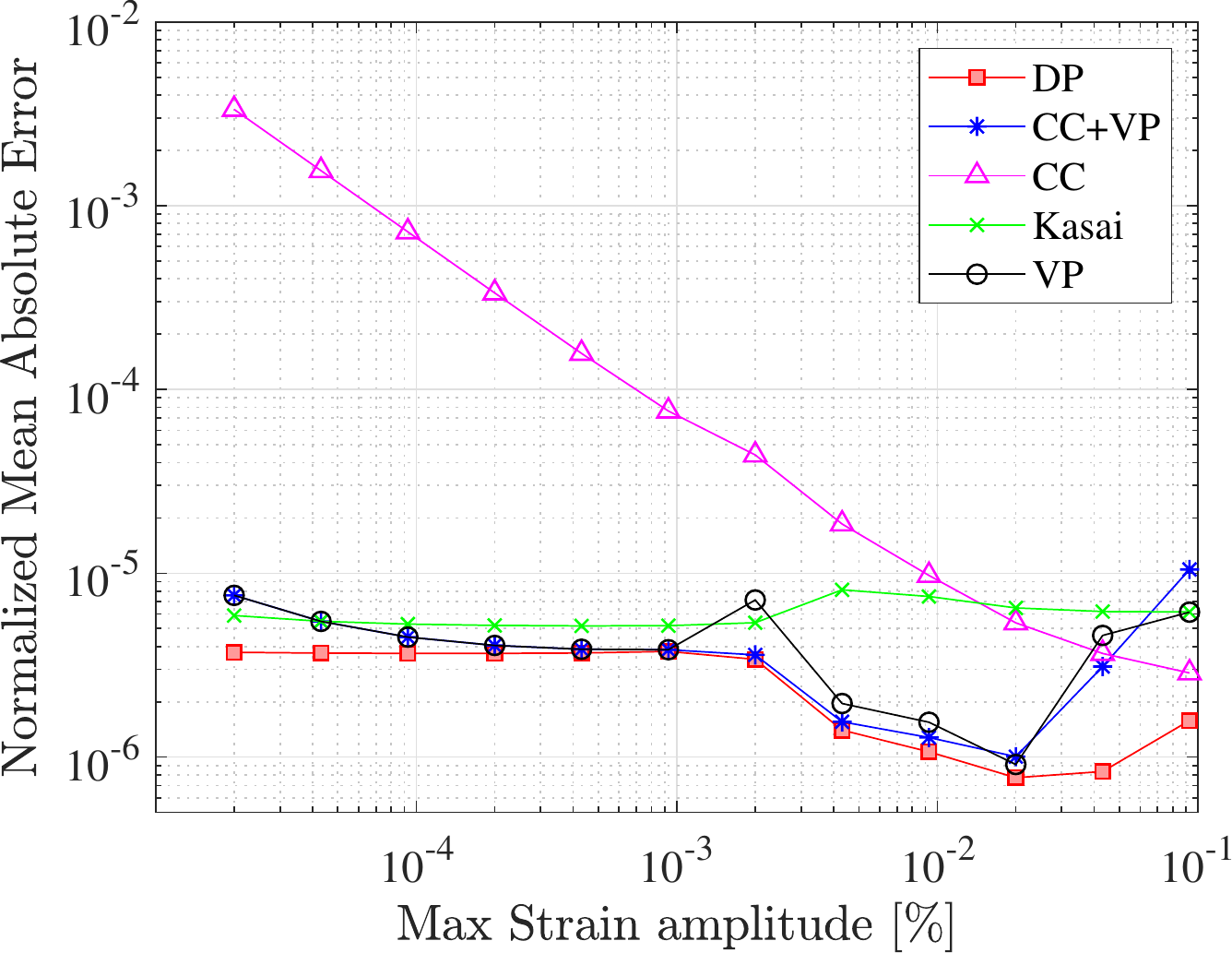}\\
		(a) Displacement errors &(b) Strain errors 
	\end{tabular}
	\caption{Normalized mean absolute errors in (a) displacements and (b) strains, by different displacement estimation algorithms for 12 simulated OCT pairs.}.
	\label{fig:simcomp}
\end{figure*}

Figures \ref{fig:simcomp}(a) and (b) show the Normalized Mean Absolute Error (NMAE) of, respectively, displacement and strain profiles for the different algorithms. 
These are computed as the absolute errors to the ground truth in Fig.~\ref{fig:profile}, averaged spatially over the frame, and then normalized to maximum strain or displacement, respectively. 

Intensity-based CC shows by far the poorest performance at low strain levels, while at higher strains showing a relative improvement (recall the normalization). 
Phase-based methods VP and Kasai expectedly perform well for small strains, but fail at higher ones due to phase wrapping. 
Since VP uses accumulated axial displacement information and it is less sensitive to phase wrapping compared to Kasai, it shows a relatively good strain estimation at higher strains, however, still rather poor performance in displacement estimation due to noise integration. 
Introducing the intensity information as well, \ie CC+VP, improves the performance of VP in the displacement estimation for larger strains. 
Our proposed DP method yields the lowest errors for both small and large displacements, confirming its advantage over the alternative approaches. 
The superiority of the DP method can be explained by a synergistic effect of combining intensity and phase information, as well as a motion continuity as a prior. 

Since both DP and VP employ the same vector-based lateral phase averaging approach for estimating phase differences for sub-wavelength displacement, they perform similarly for this range.
Nevertheless, VP includes an axial phase averaging mechanism, which may then require phase unwrapping depending on the axial window size and the strain magnitude, which can explain the superiority of DP in this range. 
For larger displacements, CC+VP outperforms VP alone by benefiting from the relative accuracy of CC algorithm in this range, while the VP algorithm additionally axially integrating the estimated strains and accounting for supra-pixel displacements. 

\section{Experimental Evaluation}
To further evaluate our method, we designed an experimental setup to examine a silicon phantom undergoing axial and lateral translations, as well as an axial compression. Displacements were induced with piezoelectric actuators, which were controlled with a low-noise high-voltage amplifier (Thorlabs MDT693B) in order to minimize the effect of electrical noise. 
A custom-built spectral-domain OCT system with a center wavelength of $877 \,\mu \mathrm{m}$ and an axial sampling precision of $4.48 \,\mu\mathrm{m}$ (in air) was used for data acquisition. All B-scans in this study were acquired with an A-scan rate of $9\, \mathrm{kHz}$, a camera integration time of $100\,\mu\mathrm{s}$ and a lateral resolution of $12 \, \mu\mathrm{m}$. 

\subsection{Axial and Lateral Translations}
\begin{figure*}
	\centering
		\includegraphics[height=.26\textwidth]{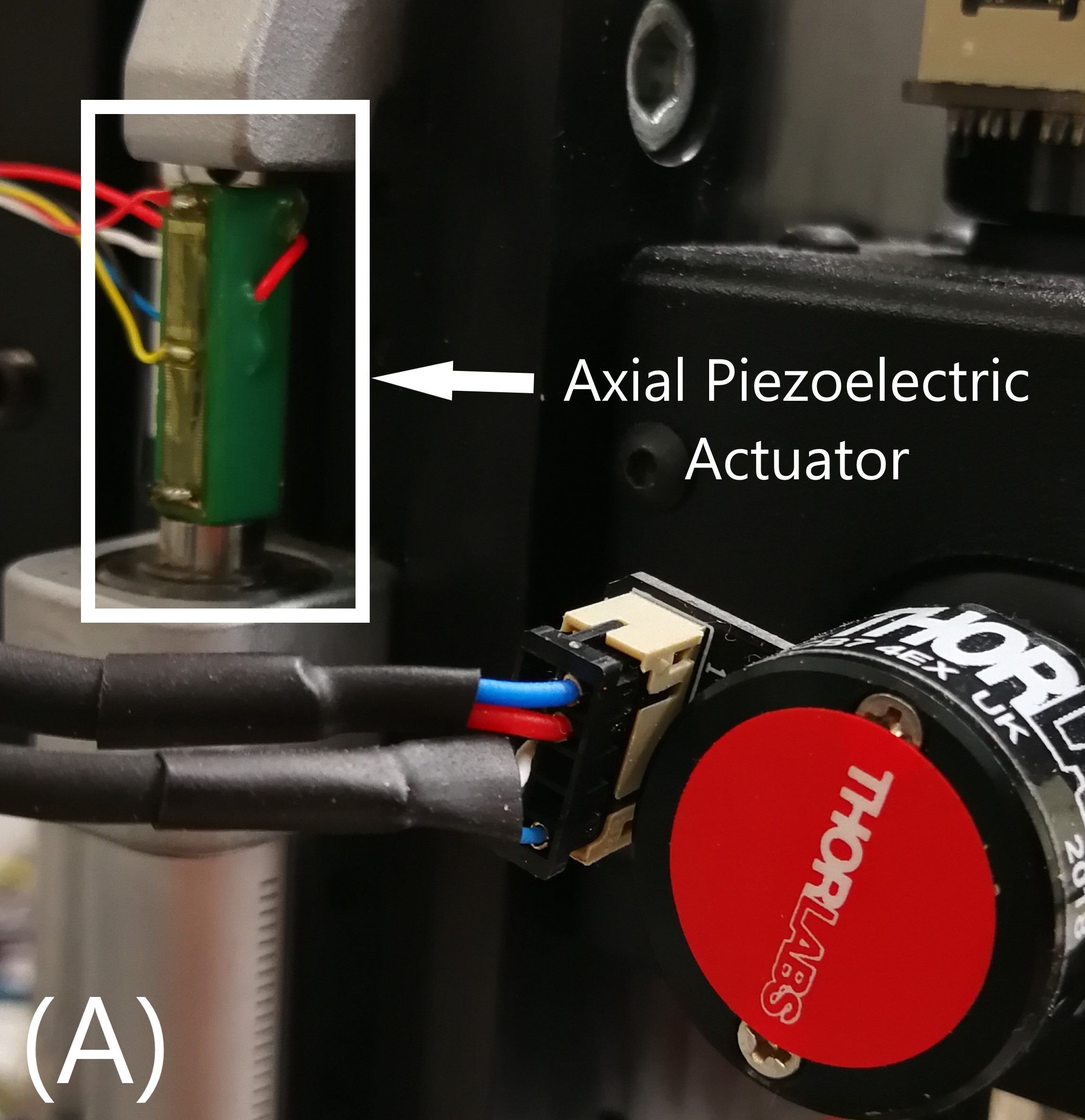}\hfill
		\includegraphics[height=.26\textwidth]{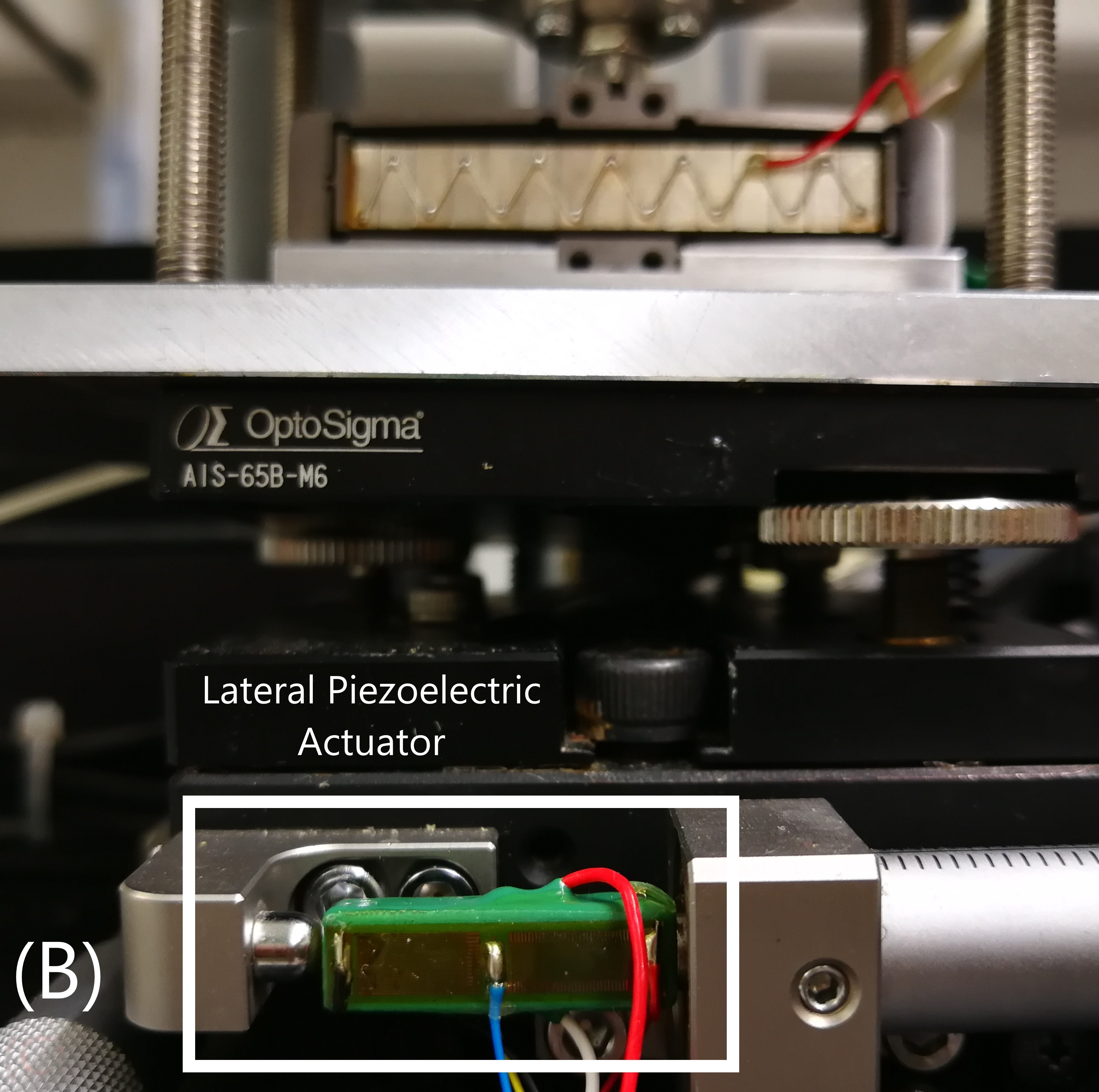}\hfill
		\includegraphics[height=.26\textwidth]{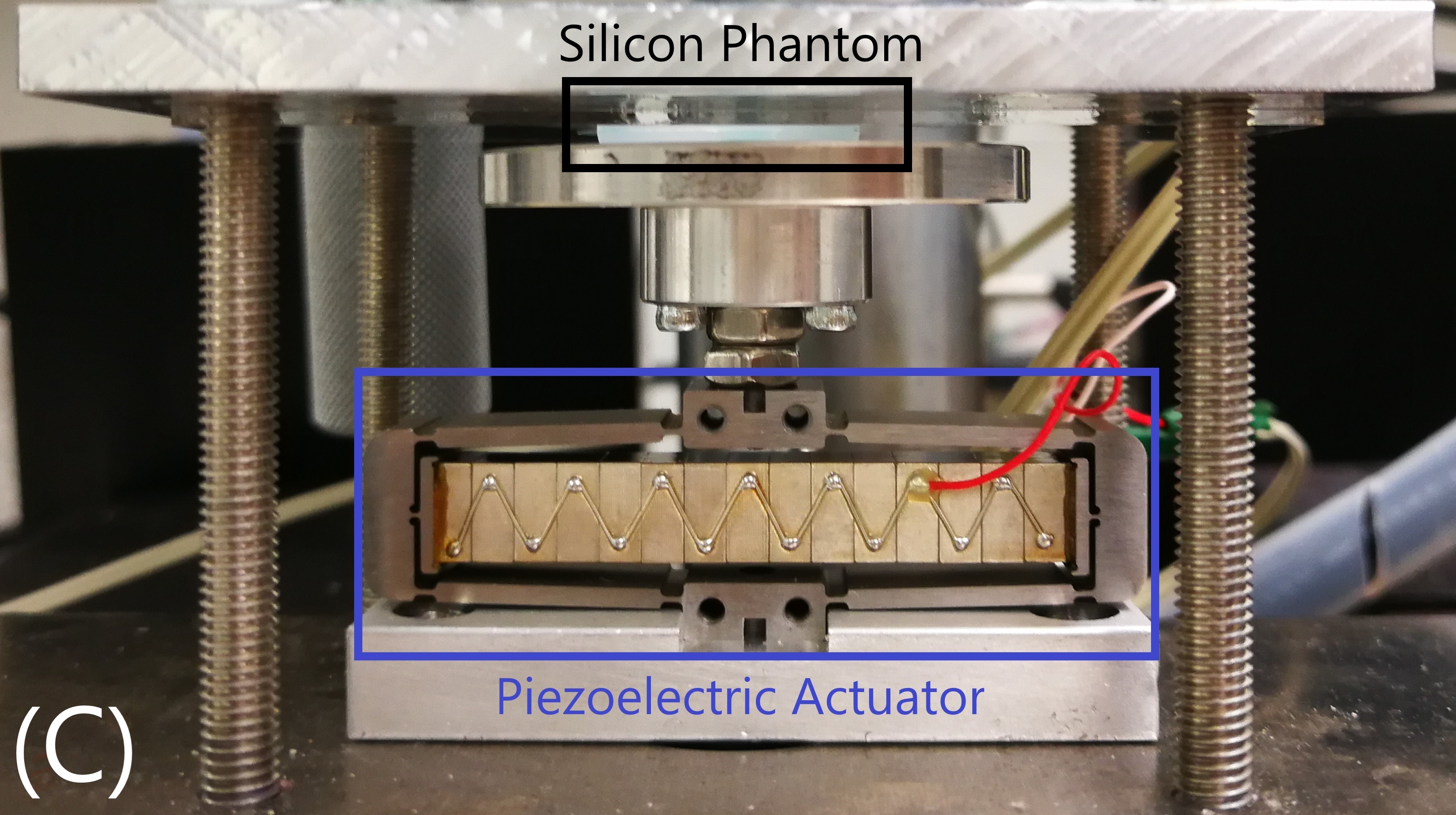}
	\caption{Experimental setup to evaluate the proposed algorithm in a silicon phantom for (a) axial and (b) lateral rigid translation, as well as (c) axial compression. Deformations were induced by piezoelectric actuators mounted in axial and lateral configurations as shown.}
	\label{fig:setup}
\end{figure*}

To assess rigid axial translation, a piezoelectric actuator (Thorlabs PZS001) was mounted on the axial motion stage of the objective lens, see Fig. \ref{fig:setup}(a). 
OCT B-scans were acquired for $13$ logarithmically-spaced rigid axial displacement steps, ranging from $1.8$\,nm to $17.32\, \mu $m. This range covers displacements from sub-wavelength scales up to about four times the pixel size, corresponding to a phase wrapping of more than 60 times. Axial displacement maps were calculated with the proposed DP method, and compared to the reference estimation algorithms mentioned above. DP parameters were set to $a^{\text{max}}=30\mu$m, $W=20$, $\beta=\gamma=10^{-5}$ and $w_1=w_2=5$. .

\begin{figure}
	\centering
		\includegraphics[width=.95\linewidth]{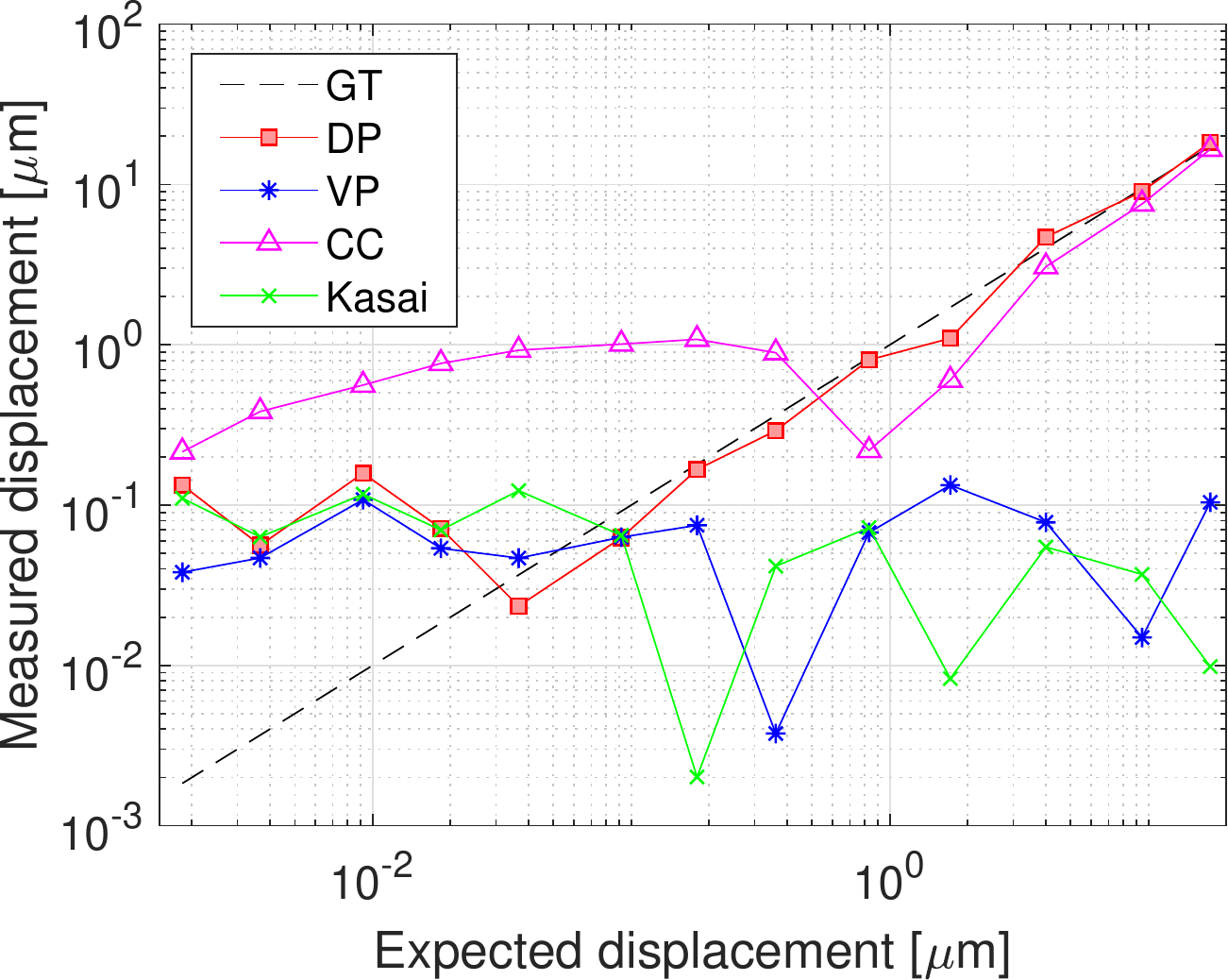}\\(a) Axial\\[1ex]
		\includegraphics[width=.95\linewidth]{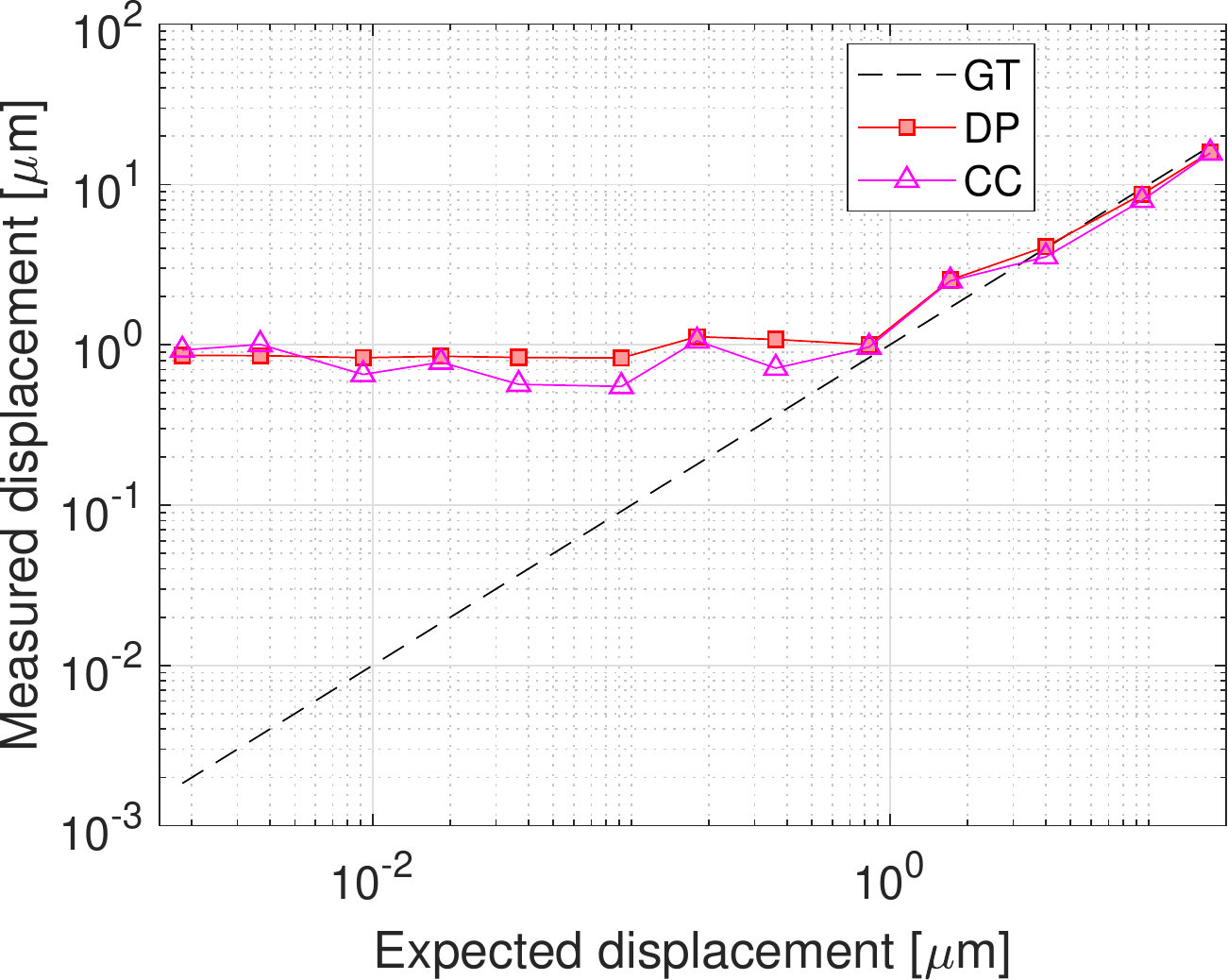}\\(b) Lateral
	\caption{Results of applying different displacement estimation algorithms to (a) axial and (b) lateral translation of the sample. Gold-truth (GT) displacements calculated based on the applied voltage and an \textit{a priori} calibration.}.
	\label{fig:axial}
\end{figure}

Fig.~\ref{fig:axial}(a) shows the axial displacements estimated by the different algorithms. Our proposed DP method outperforms the other methods for both small and large displacements. For displacements below $0.04\mu$m corresponding spatially to the phase noise level of the OCT setup, all of the algorithms fail to estimate the displacements. 

To assess lateral translation, a piezoelectric actuator was mounted on the lateral motion stage of the sample, see Fig. \ref{fig:setup}(b). Similarly as above, OCT B-scans were acquired for $13$ logarithmically-spaced rigid lateral displacement steps, ranging from $1.8\,$nm to $17.32\,\mu$m. Here, we compared DP only to the CC method, given that phase-based methods are unable to provide information on lateral displacements. 
Fig.~\ref{fig:axial}(b) shows the estimated lateral displacements of both these algorithms, indicating similar performance by both methods, especially above the noise level. 
This shows that the substantially higher axial accuracy of DP does not come as a tradeoff for lateral accuracy.

\subsection{Axial Compression: OCE}
To assess axial compression, the silicon phantom was placed between a microscope slide and a piezoelectric actuator (Thorlabs APF705), as shown in Fig. \ref{fig:setup}(c). Since the top/bottom surfaces of our silicon phantom were stuck on the above/below surfaces, they are assumed to be fixed, while the side surfaces are assume to be free. This means that we expect to see the maximum axial and lateral strains in the middle depth of the phantom.  Again, OCT B-scans were acquired for $12$ logarithmically-spaced compression amplitudes created by applying voltages between $0$ (zero displacement) to $75\,V$, which corresponds to about 220\,$\mu$m displacement when the piezoelectric actuator is not under any load.
The investigated displacement range covers sub-wavelength to supra-pixel scales. 
Nevertheless, since the actuator stroke changed significantly based on the load on it, the exact applied displacements and hence the ground truth were not unknown precisely. 
Therefore, the measurement settings (x-axes) are expressed in terms of applied voltage, rather than corresponding displacements.

\begin{figure*}
	\centering
	\begin{tabular}{@{}ccc@{}}
		\includegraphics[height=0.245\textwidth]{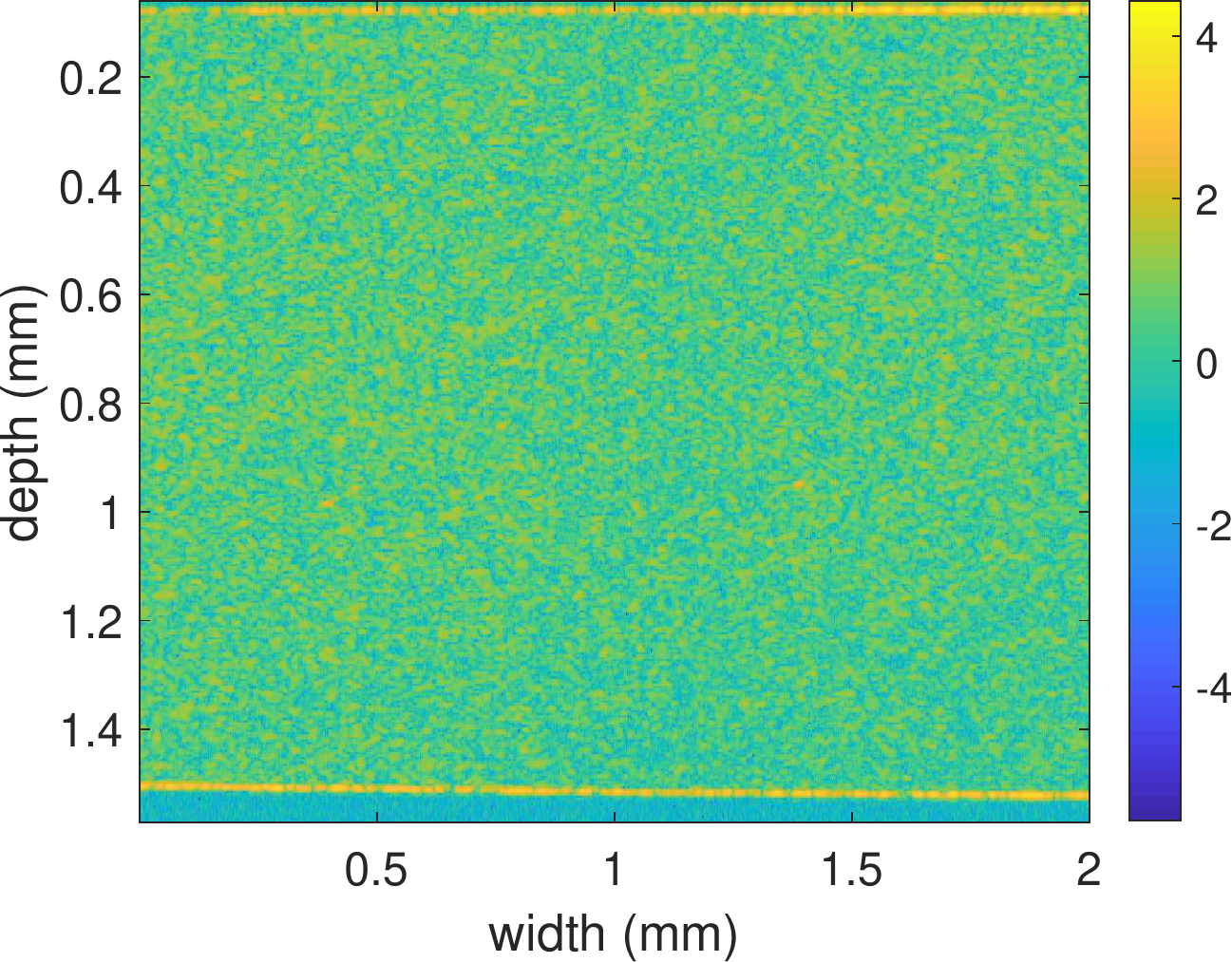}&
		\includegraphics[height=0.245\textwidth]{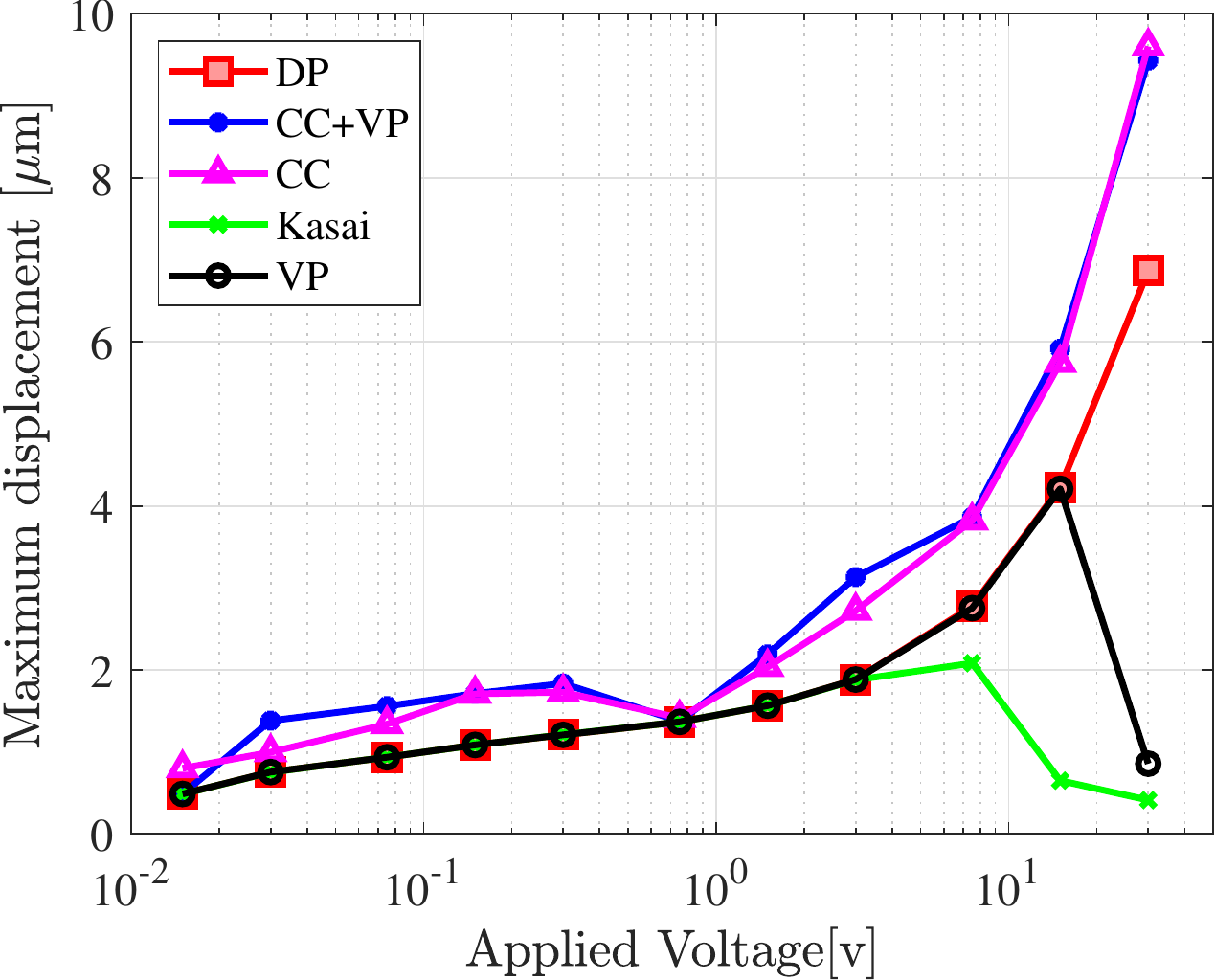}&
		\includegraphics[height=0.245\textwidth]{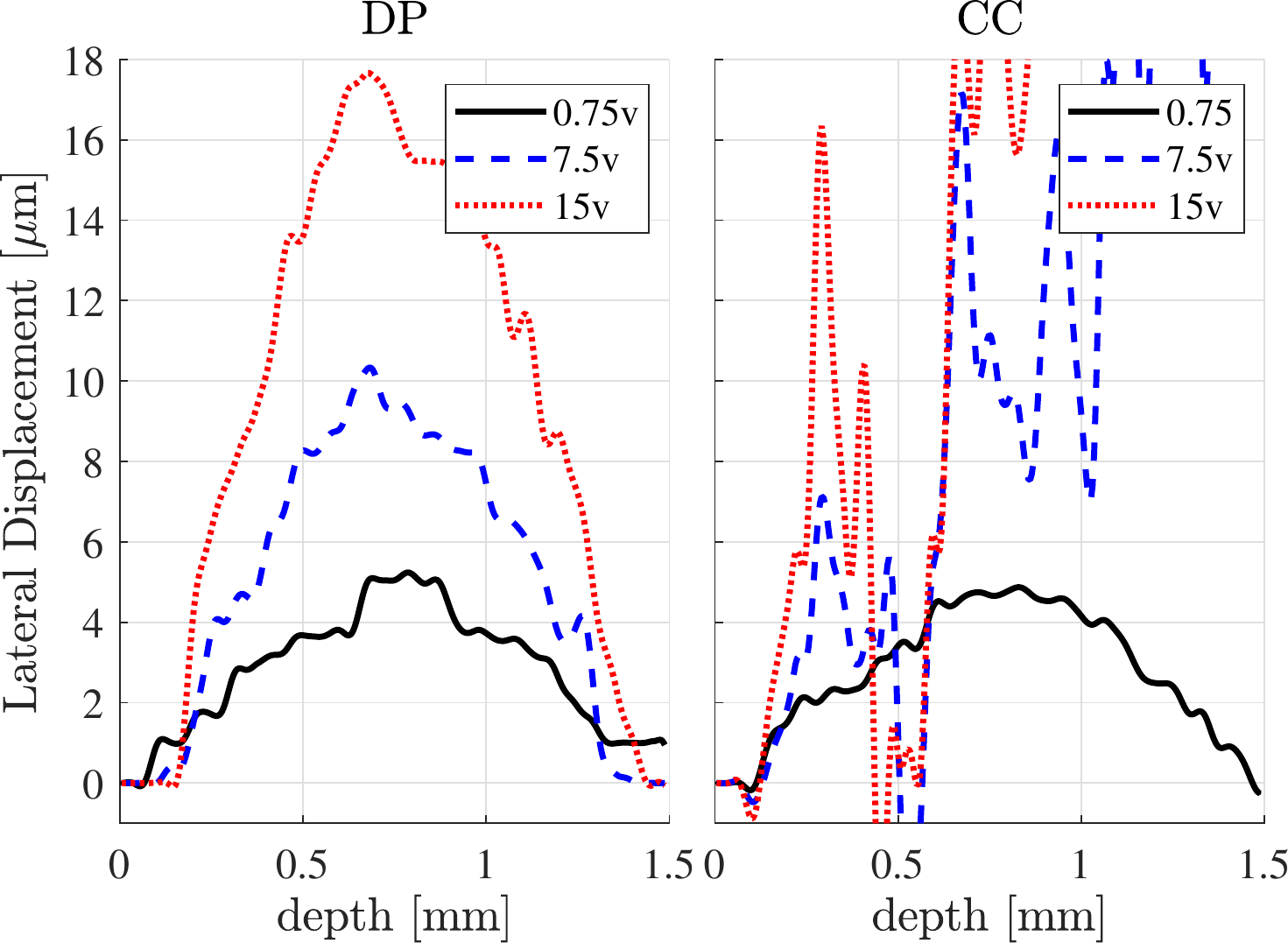}\\
		(a)&(b)&(c)
	\end{tabular}
	\caption{(a) The pre-deformed OCT signal magnitude, acquired such that the maximum displacement measured at the end of each axial line is of the piezoelectric actuator. (b)~Comparison of displacement estimation methods, based on maximum estimated displacement for voltage applied on the actuator.  (c)~shows the the estimated lateral displacements by DP and CC methods averaged over all axial lines for 0.75v, 7.5v and 15v experiments.}
	\label{fig:compressional}
\end{figure*}

Fig.~\ref{fig:compressional}(a) shows a typical OCT signal magnitude image in this experiment, in which signal decorrelation due to speckle boiling/blinking is significant for large displacements. 
Note that due to the non-contact nature of OCT one can not make sure that the surface displacements are zero. Herein, the top of the phantom was fixed with a microscope slide, which however bends slightly under piezoelectric actuator load (see Fig.~\ref{fig:setup}(c)).  
This causes the sample not only compress from one side, but also translate altogether, which is one of the reasons for non-available ground-truth measurements.
Furthermore, although CC and DP algorithms inherently compensate for such translations, we observed any purely phase-based method to fail completely when such translation was not compensated.
Thus, to have a fair comparison for this compression experiment, we compensated the translation for Kasai and VP methods, using the average CC estimate in an axial window of $48\mu$m at a depth of $80\mu$m, where the CC estimations were observed to be stable. 
DP parameters were set to $a^{\text{max}}=500\mu$m, $W=20$, $\beta=0.5$, $\gamma=0.5$, $l^{\text{max}}=20\mu$m, $w_1=w_2=10$ and lateral sub-pixel resolution of $0.5 \mu$m. 

Although the GT displacements are unknown, it is a safe assumption to expect increasing (maximum) displacements, with increasing actuator voltages.
Figure \ref{fig:compressional}(b) shows the maximum displacements estimated by different algorithms.
The values are calculated using the median value in an axial window of $48\,\mu$m at a depth of $1.32$\,mm minus the median value in an axial window of $48\,\mu$m at a depth of $80\,\mu$m over all axial lines. 
DP is the only method that exhibits monotonically increasing maximum displacements with increasing voltages.
Furthermore, for small displacements DP is in good agreement with Kasai and VP, which are the phase-based methods that are known to work reliably at small displacement regimes.
Among these three, only DP performs also well at larger displacements, \ie continues with reasonable monotonically increasing estimates. 

For three voltage settings, Figure \ref{fig:compressional}(c) shows the average lateral displacement estimation by DP and CC, the only two methods that estimate lateral motion herein.
Although we do not have ground-truth lateral measurements either, for such a homogenous sample and uniaxial compression profile, it is quite reasonable to expect smooth displacement fields, which is used herein as a qualitative criterion to assess the lateral estimations.
As seen in the figures, DP presents relatively smooth axial profiles (of average lateral displacements), which also start and end around zero, irrespective of the voltage setting.
In contrast, CC for the voltage settings 7.5\,V and 15\,V exhibit totally erratic behaviour that is not expected from a physical sample, while also not returning to zero at higher depth (probably due to accumulating errors).

\begin{figure*}
	\centering
	\begin{tabular}{@{}ccc@{}}
		\includegraphics[height=0.265\textwidth,clip,trim=1.5em 0 3em 0]{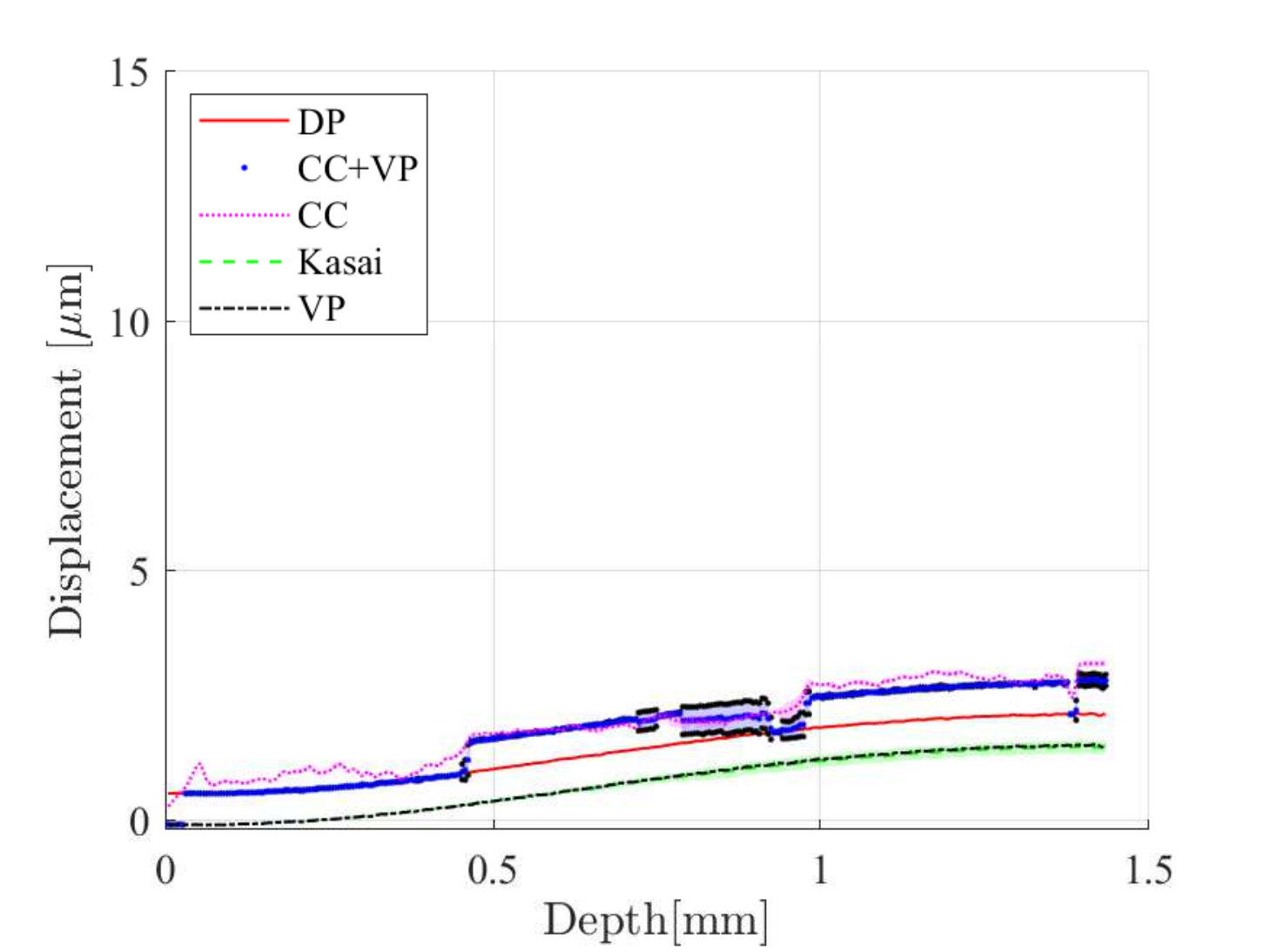}&
		\includegraphics[height=0.265\textwidth,clip,trim=1.5em 0 3em 0]{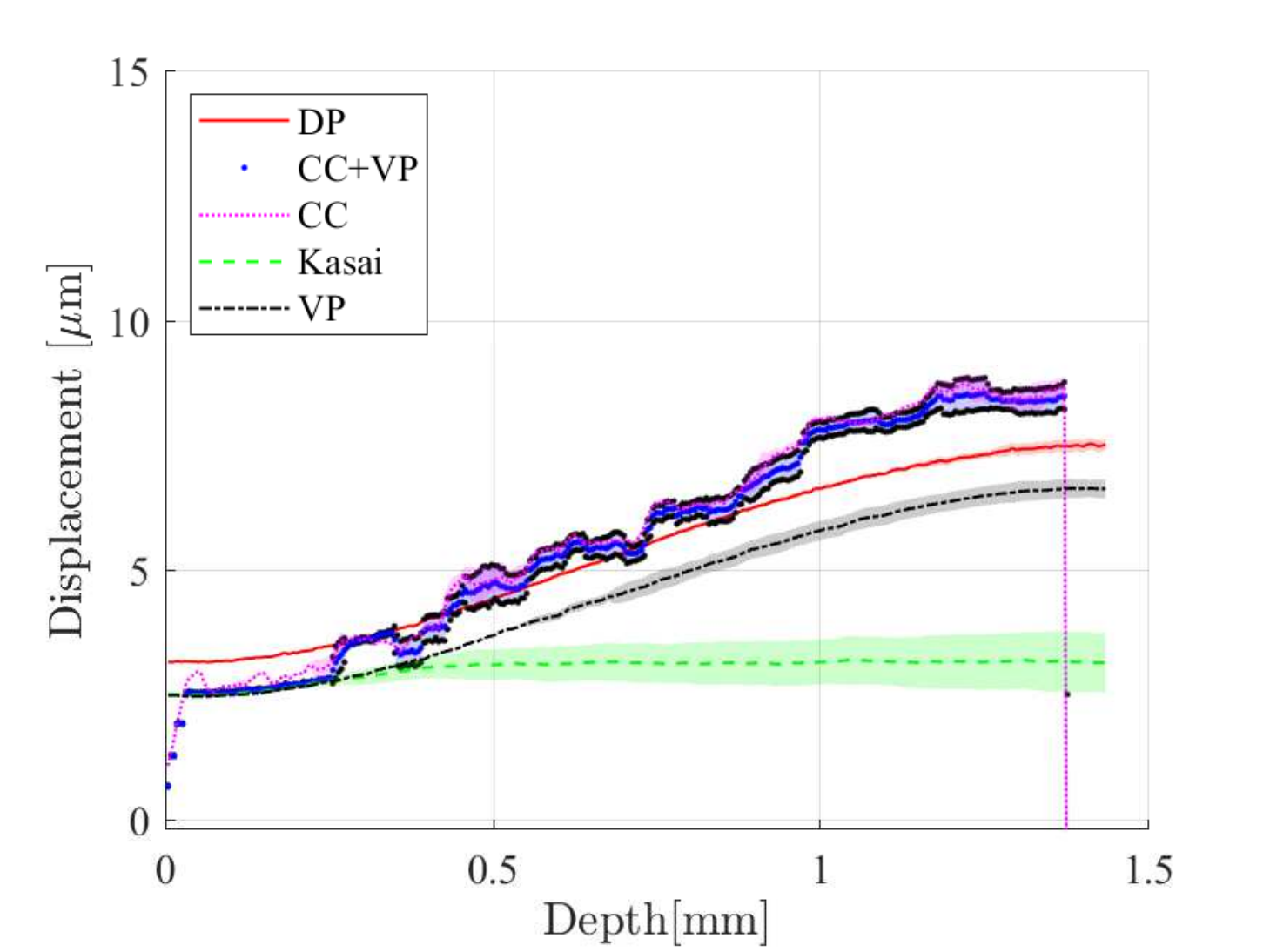}&
		\includegraphics[height=0.265\textwidth,clip,trim=1.5em 0 3em 0]{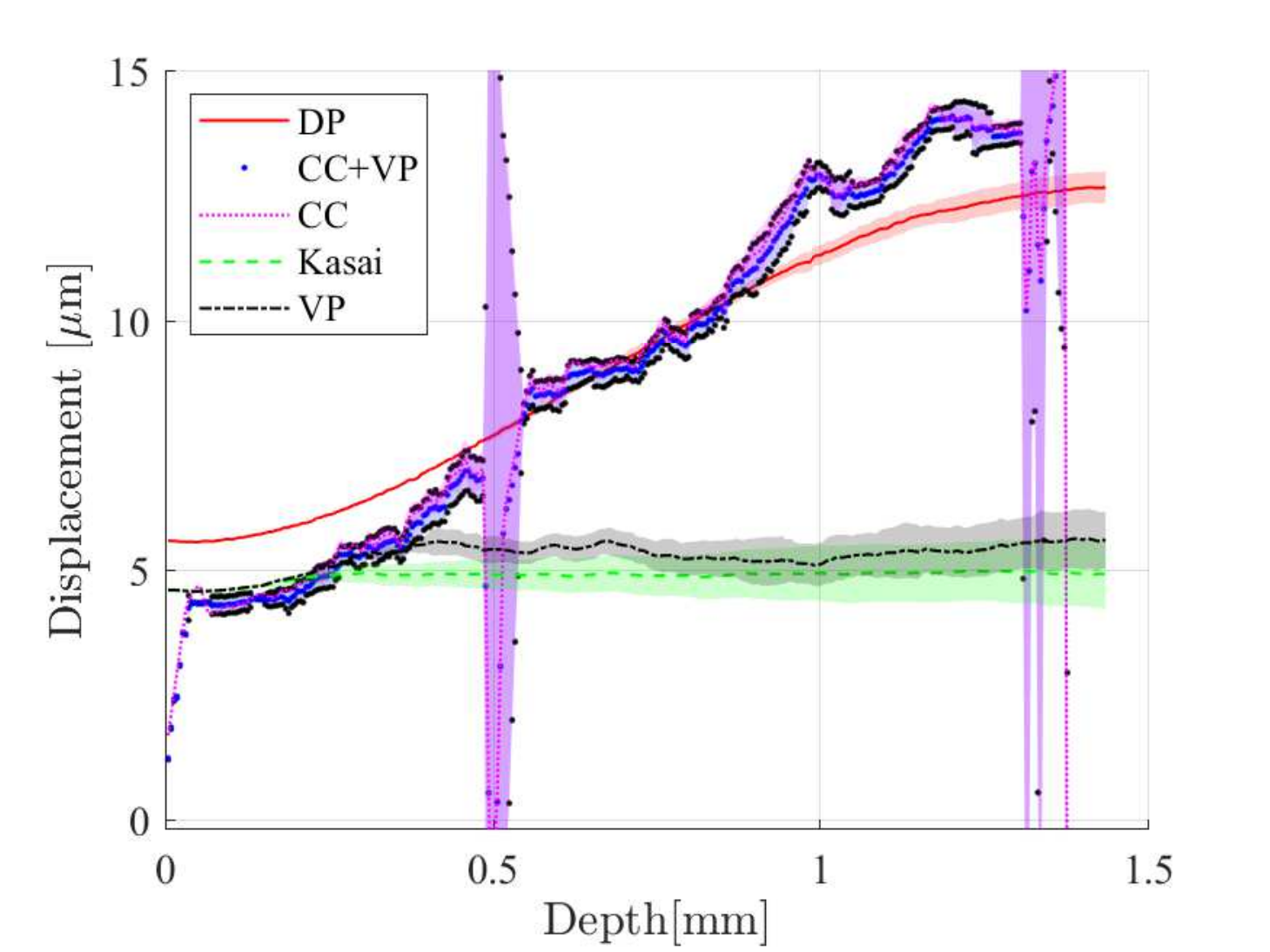}\\
		(a) Displacements for 0.75\,V&(b) Displacements for 7.5\,V&(c) Displacements for 15\,V\\
		\includegraphics[height=0.265\textwidth,clip,trim=1em 0 3em 0]{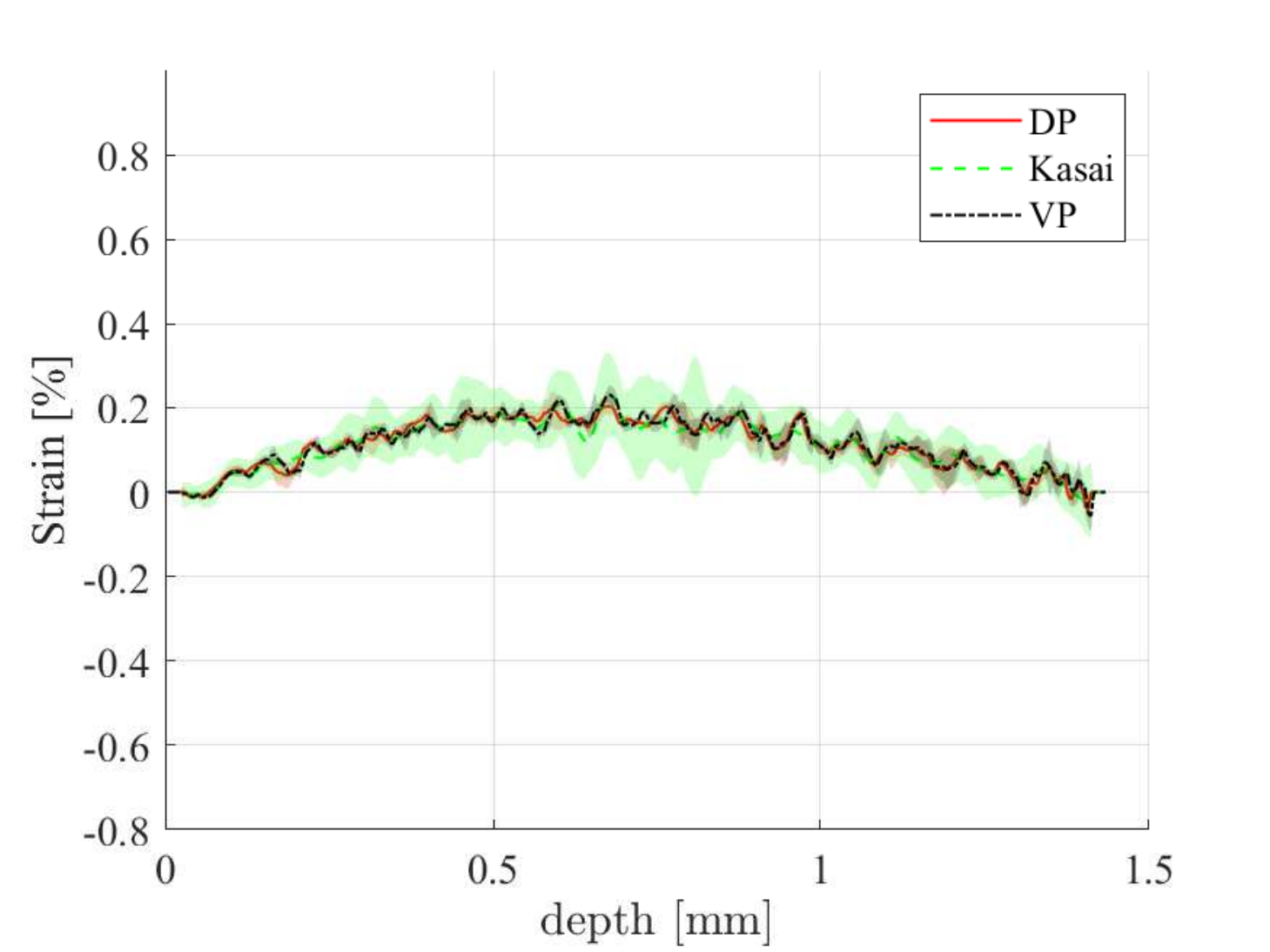}&
		\includegraphics[height=0.265\textwidth,clip,trim=1em 0 3em 0]{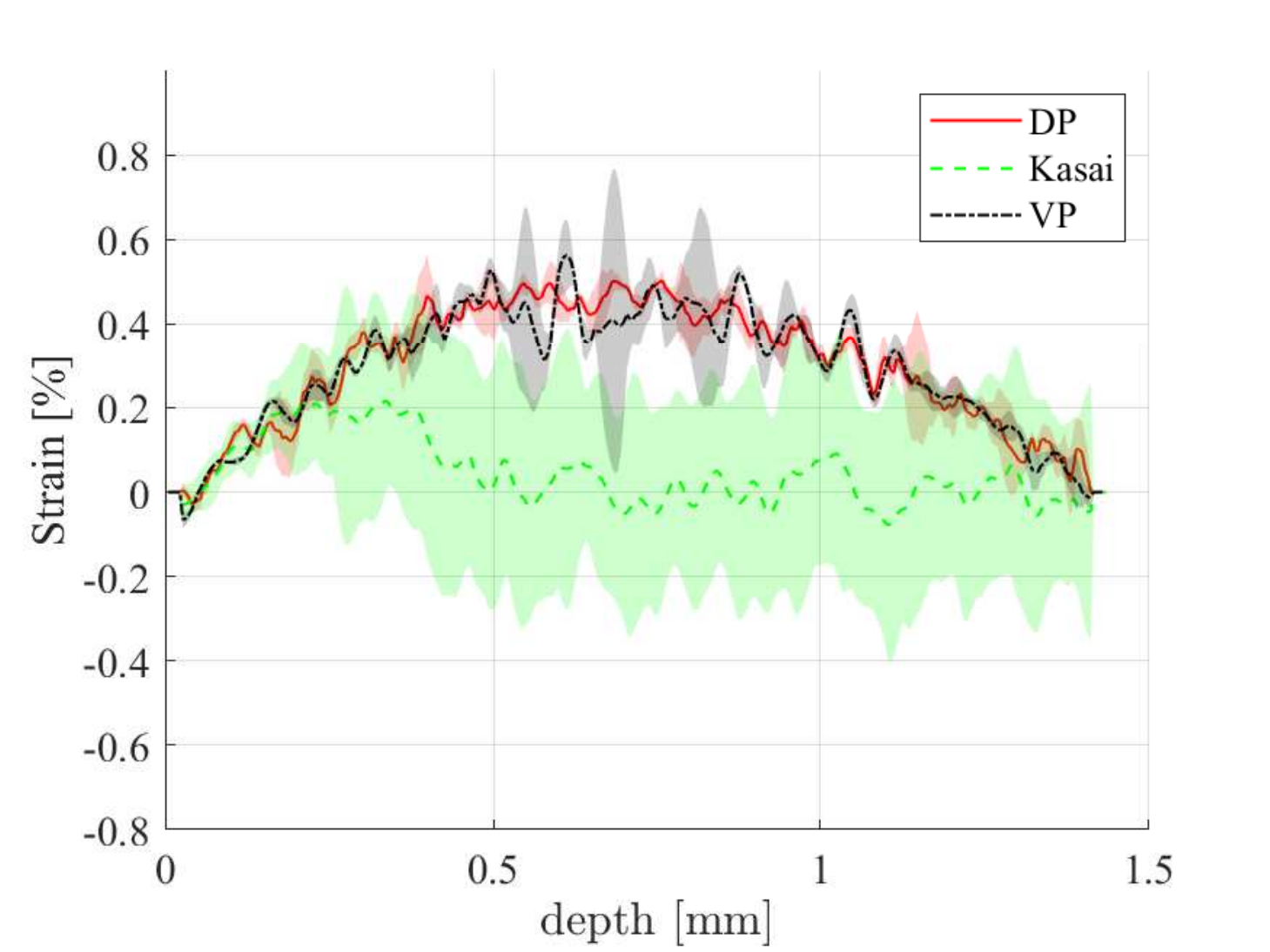}&
		\includegraphics[height=0.265\textwidth,clip,trim=1em 0 3em 0]{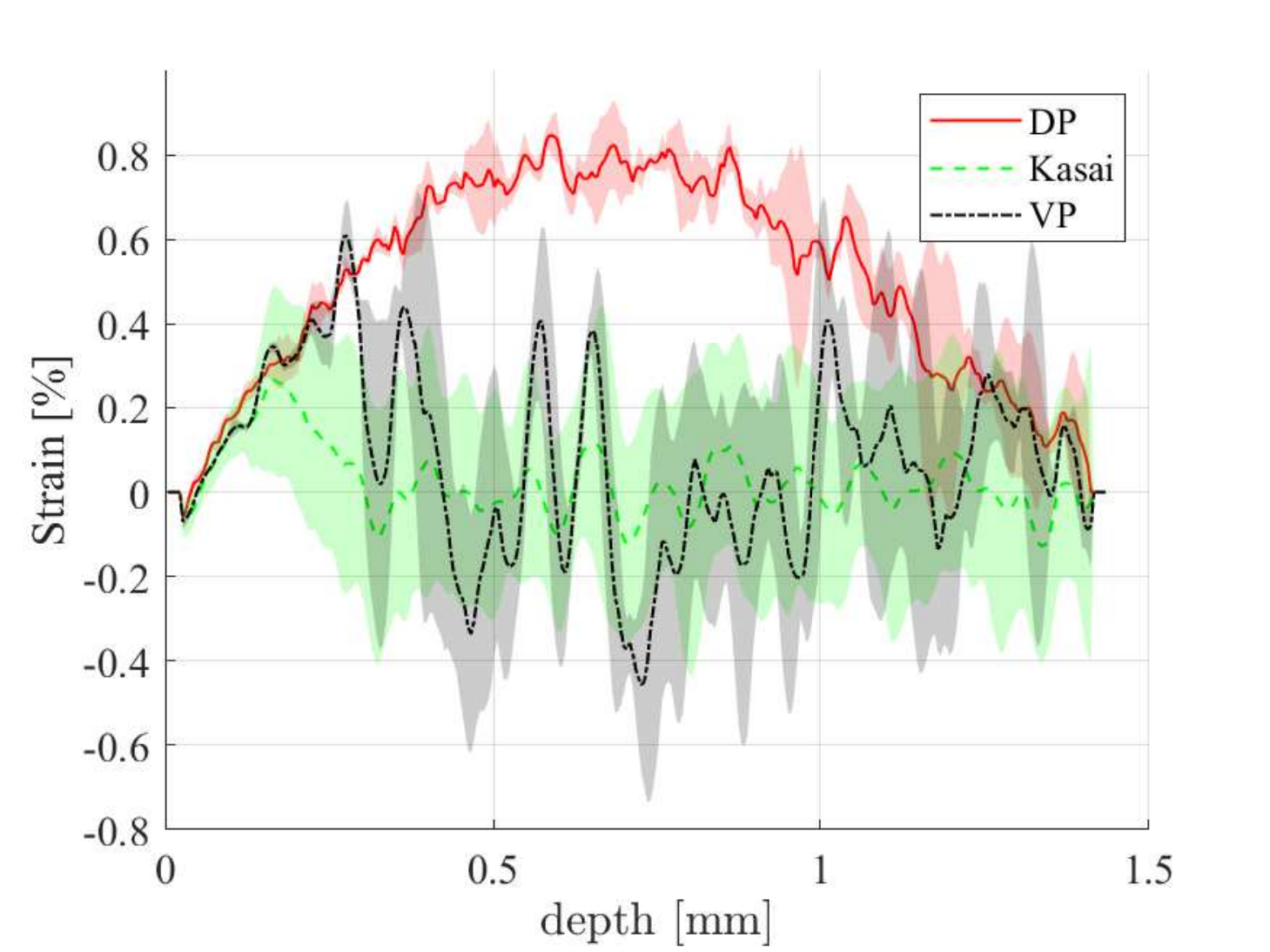}\\
		(d) Strains for 0.75\,V&(e) Strains for 7.5\,V&(f) Strains for 15\,V
	\end{tabular}
	\caption{Comparison of axial (a)-(c) displacement and (d)-(f) strain profiles of algorithm for 0.75\,V, 7.5\,V, and 15\,V experiments. Per-depth mean (lines) and standard deviation (shaded region) of axial estimates for 100 A-lines across the phantom are shown.}
	\label{fig:compressional_example}
\end{figure*}

For axial displacements, in addition to comparing the maximum estimated values as above, we also evaluate in Fig.~\ref{fig:compressional_example} the axial estimation profiles from different methods for three experimental settings: 0.75\,V, 7.5\,V, and 15\,V (chosen at critical values where each a different method is suspected to fail).  
For 100 lateral estimation profiles (A-lines) across the sample, we plot in the figure the mean and standard deviations of axial displacement estimations per depth.
Fig.~\ref{fig:compressional_example} shows that at 0.75\,V 
DP, VP, and Kasai are all successful in finding the displacement/strain profiles (although Kasai with larger deviations in strain), whereas CC and CC+VP methods fail due to the poor estimates of the CC algorithm. 
At 7.5\,V, the maximum displacements reach supra-pixel values and thus the Kasai algorithm fails while VP also starting to show signs of false jumps (seen as amplified in the strain profiles). 
At 15\,V only DP can estimate displacements and strains with low standard deviation, also yielding reasonable profiles.

\begin{figure*}
	\centering
	\begin{tabular}{ccc}
		\includegraphics[height=0.25\textwidth]{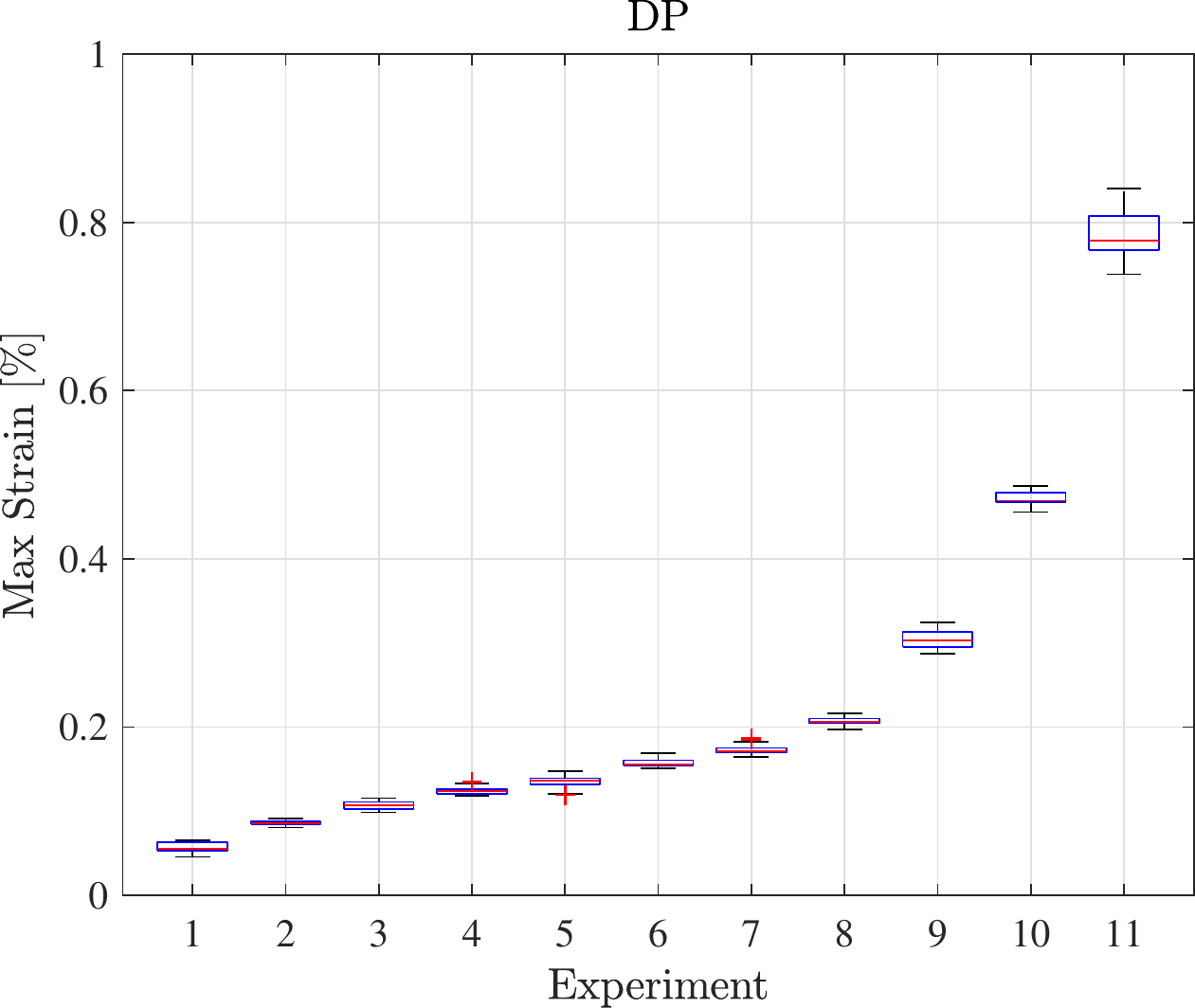}&\includegraphics[height=0.25\textwidth]{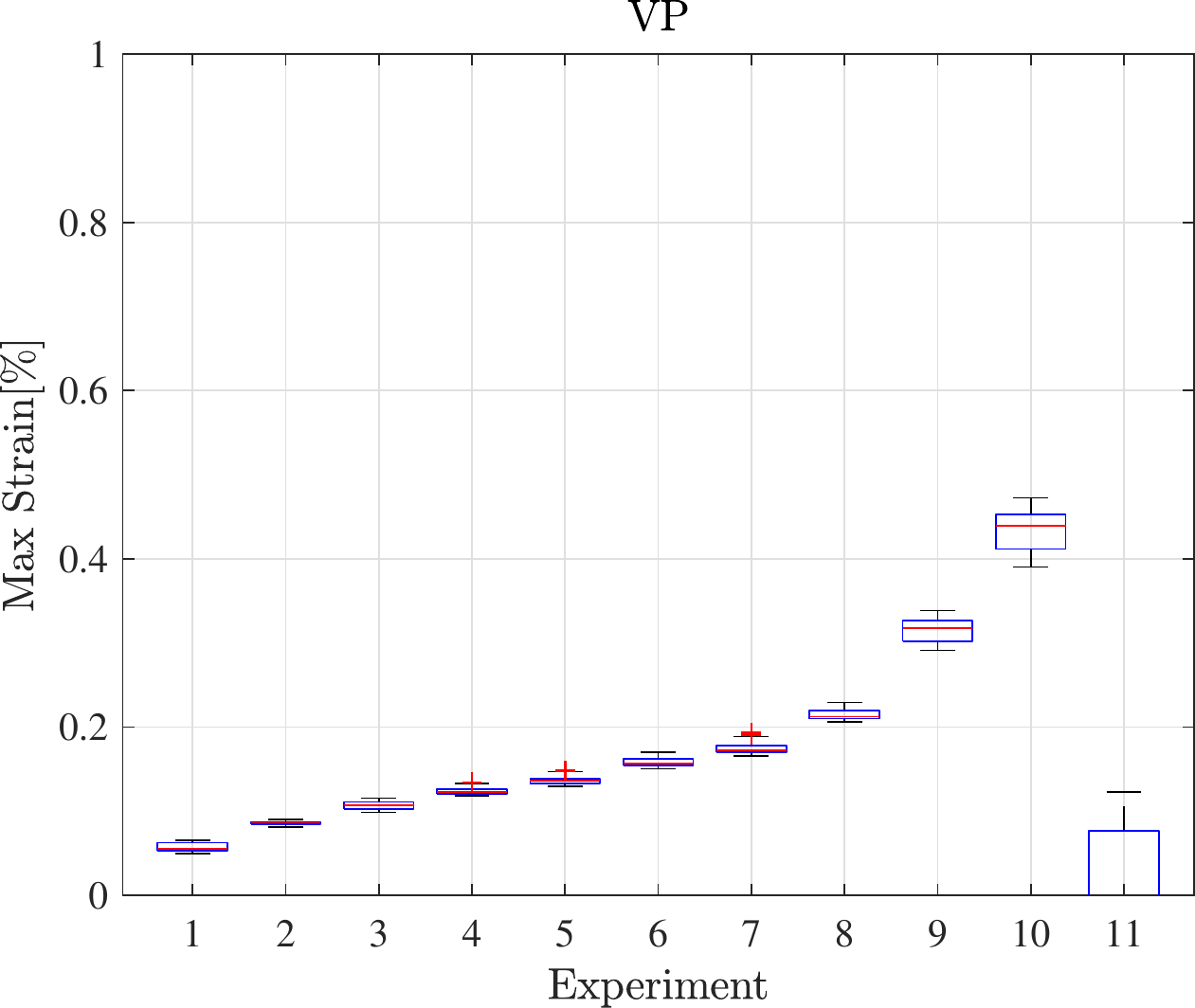}&\includegraphics[height=0.25\textwidth]{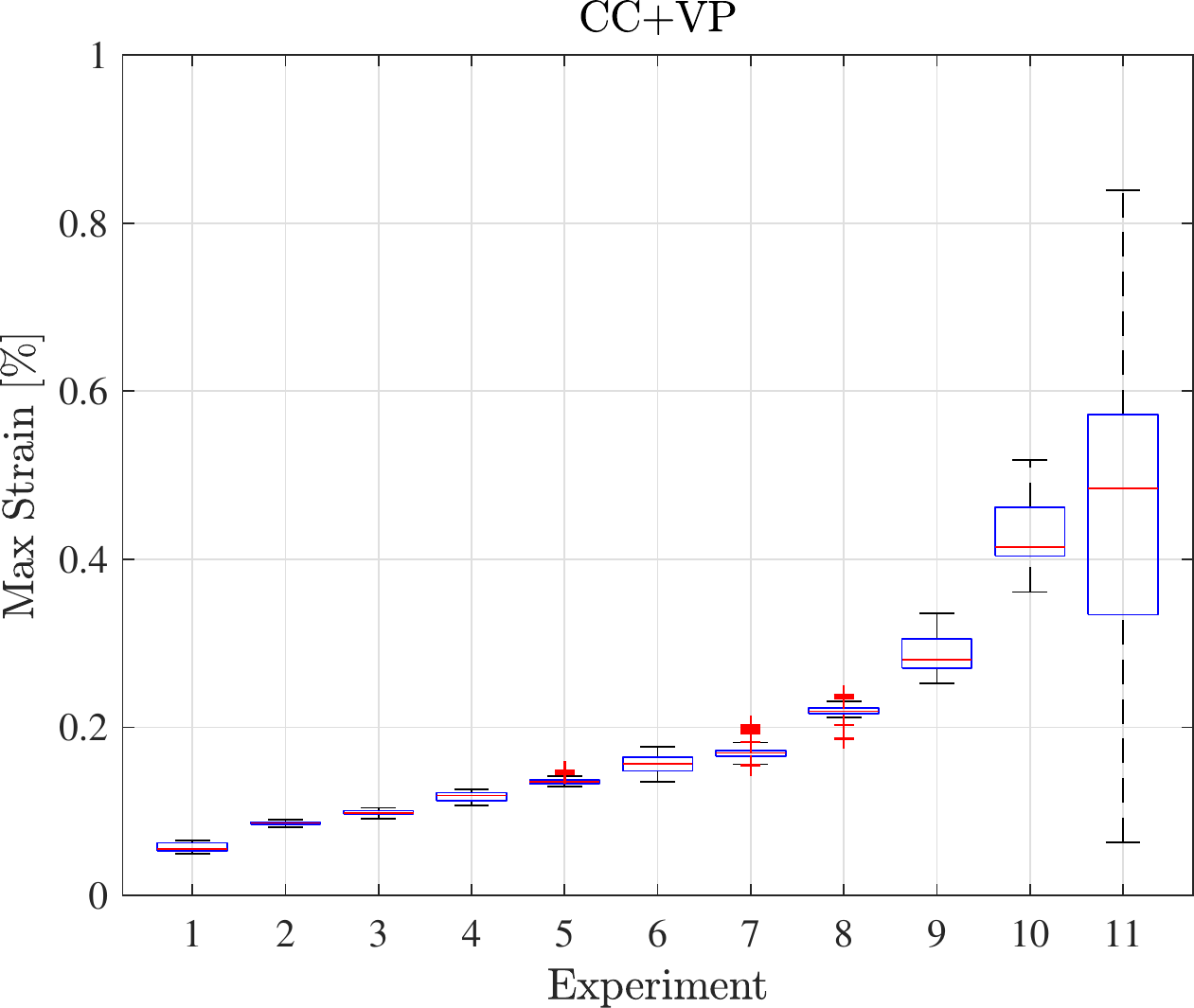}\\
		(a)&(b)&(c)\\
		\includegraphics[height=0.25\textwidth]{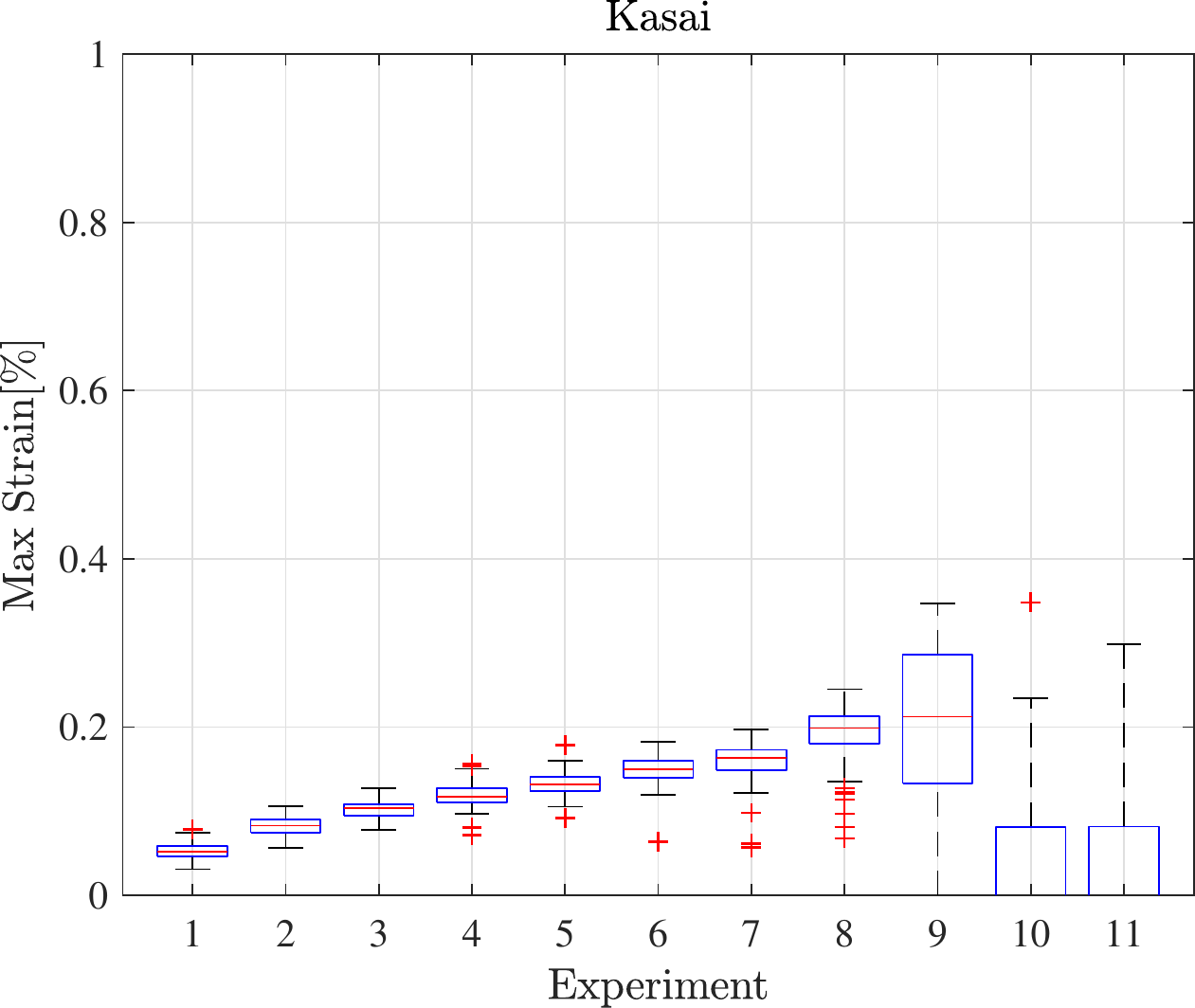}&\includegraphics[height=0.25\textwidth]{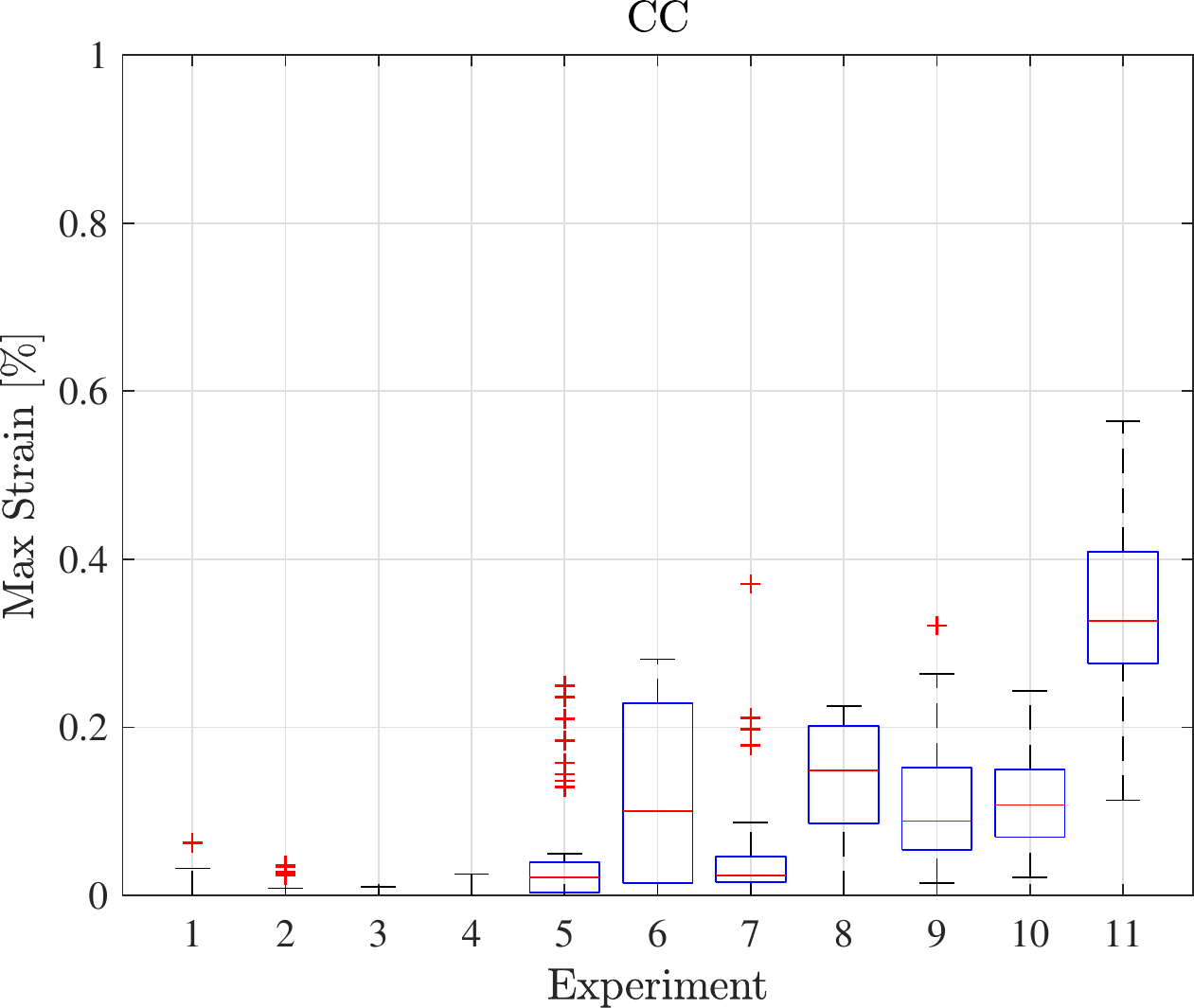}&\includegraphics[height=0.25\textwidth]{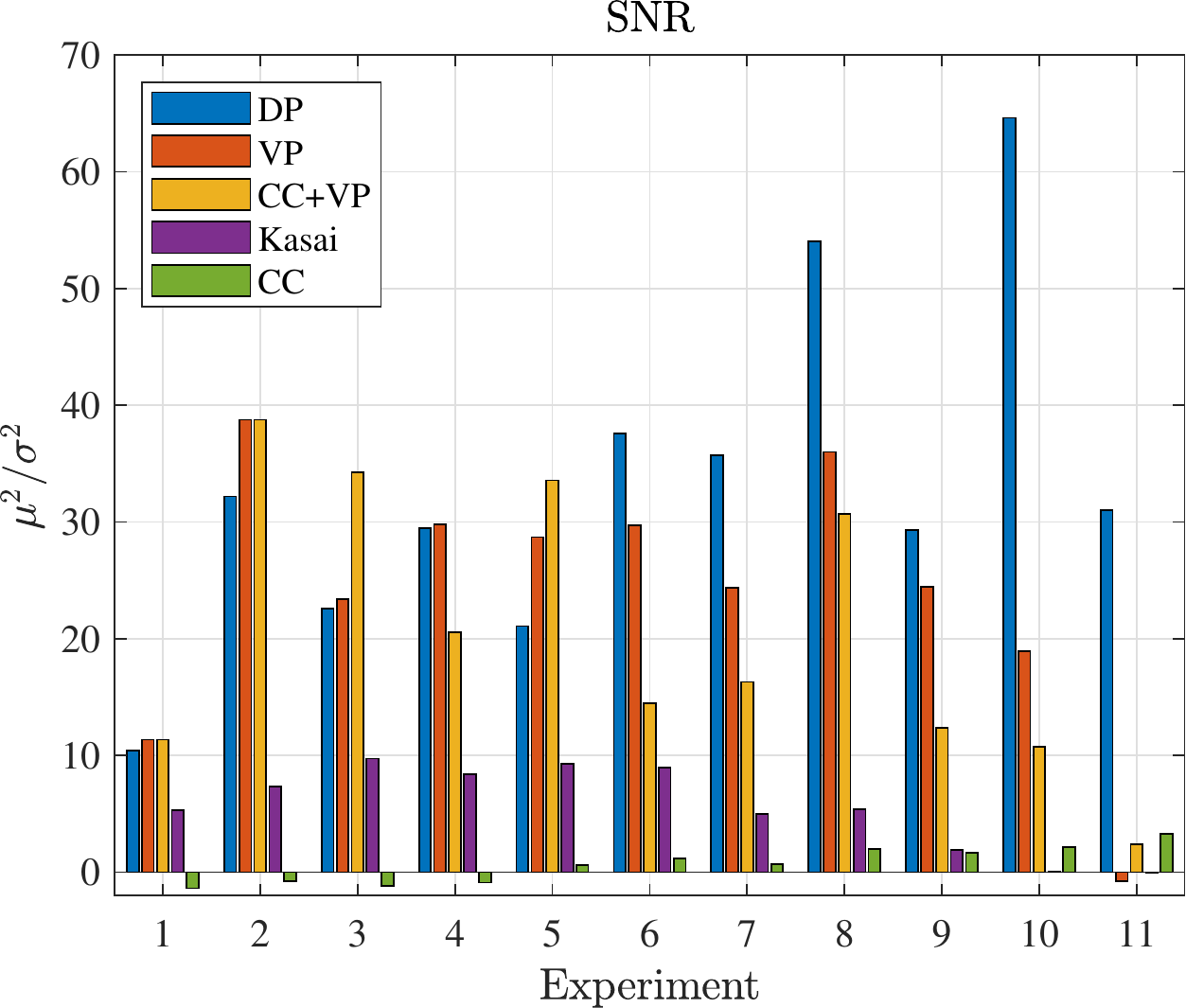}\\
		(d)&(e)&(f)
	\end{tabular}
	\caption{Distributions of max axial strain value within a phantom depth of [0.6, 0.8]\,mm across different estimation A-lines, where a larger variation indicates lower SNR, given the relative homogeneity of the phantom and the compression.
	Results reported for the 11 compression magnitudes for the five algorithms in~(a)-(e). In the box-plots, the central mark indicates the median, with the bottom and top edges indicating the 25th and 75th percentiles, respectively. The whiskers extend to the most extreme data points not considered as outliers, which are plotted individually as '+'. 
	(f)~shows the SNR of the different algorithms in each of the 11 experiments}
	\label{fig:compressional_boxplot}
\end{figure*}

Fig.~\ref{fig:compressional_boxplot} shows a comparison of the different algorithms in estimating the maximum axial strain, at a phantom depth of $\approx$0.7\,mm for the 11 compression magnitudes.
Given the variations, Fig.\,\ref{fig:compressional_boxplot}(f) reports the SNR per compression magnitude and for each algorithm, indicating that DP achieves optimal SNR for strain estimations, especially for higher displacement magnitudes. 
With the proposed method yielding high SNR and high quality strain/displacement estimates in a much wider deformation range can alleviate data acquisition limitations in OCE.

%

\section{Conclusions}
\label{sec:Conc}

We have introduced a robust displacement estimation method, which can efficiently track axial and lateral displacements in OCT with, respectively, sub-wavelength and pixel scale resolutions.
The displacement tracking is formulated as an optimization problem solved using dynamic programming, which utilizes both the sub-wavelength-scale phase and pixel-scale intensity information of OCT B-scans to solve for axial and lateral displacements. 
The DP cost function is designed to find the number of phase wraps, in addition to the measured phase difference between a reference and deformed OCT scans, by minimizing the motion discontinuity and intensity disparity. 
This proposed method has several advantages: 
First, it introduces the benefits of adding the motion continuity as a prior information in OCE. 
Second, it estimates pixel-scale and sub-wavelength-scale displacements at the same time as opposed to conventional methods in the literature that use either the intensity information only to estimate large pixel-scale displacements or the phase information only to calculate sub-wavelength displacements. 
These earlier approaches are prone to decorrelation caused by speckle blinking and boiling. 
Our experimental results on a silicon phantom showed that the proposed DP method is able to efficiently estimate displacement maps from quasi-static compressions, substantially outperforming existing methods in terms of axial tracking precision for displacements exceeding half the central wavelength of the OCT beam. 
In addition, as shown in the axial compression tests, deformation estimation was feasible on a single A-line without lateral averaging. 
Hence, our proposed DP method is expected to be able to track dynamic OCE signals as well as motion in M-scans with similar precision, a feature not easily possible with the conventional methods. 
A potential limitation of our method is its high computational complexity; nevertheless, the maximum computation time for the results presented herein was approximately a minute on a quad-core CPU. 
Furthermore, several extensions of the DP algorithm have been proposed in the literature for reducing its computational cost.
We believe that our proposed method yielding high quality displacement estimates for a wide range of deformations will greatly simplify and foster OCE, in particular, and potentially other OCT-based methods in general.

\section*{Acknowledgment}

This study was funded by the Swiss National Science Foundation (Ambizione grant PZ00P2\_174113 of SK).

\ifCLASSOPTIONcaptionsoff
  \newpage
\fi

\bibliographystyle{IEEEtran}
\bibliography{Main}

\begin{thebibliography}{10}
\providecommand{\url}[1]{#1}
\csname url@samestyle\endcsname
\providecommand{\newblock}{\relax}
\providecommand{\bibinfo}[2]{#2}
\providecommand{\BIBentrySTDinterwordspacing}{\spaceskip=0pt\relax}
\providecommand{\BIBentryALTinterwordstretchfactor}{4}
\providecommand{\BIBentryALTinterwordspacing}{\spaceskip=\fontdimen2\font plus
\BIBentryALTinterwordstretchfactor\fontdimen3\font minus
  \fontdimen4\font\relax}
\providecommand{\BIBforeignlanguage}[2]{{%
\expandafter\ifx\csname l@#1\endcsname\relax
\typeout{** WARNING: IEEEtran.bst: No hyphenation pattern has been}%
\typeout{** loaded for the language `#1'. Using the pattern for}%
\typeout{** the default language instead.}%
\else
\language=\csname l@#1\endcsname
\fi
#2}}
\providecommand{\BIBdecl}{\relax}
\BIBdecl

\bibitem{kennedy2017emergence}
B.~F. Kennedy, P.~Wijesinghe, and D.~D. Sampson, ``The emergence of optical
  elastography in biomedicine,'' \emph{Nature Photonics}, vol.~11, no.~4, p.
  215, 2017.

\bibitem{Larin2017}
K.~V. Larin and D.~D. Sampson, ``Optical coherence elastography ; oct at work
  in tissue biomechanics,'' \emph{Biomedical {O}ptical {E}xpress}, vol.~8,
  no.~2, pp. 1172--1202, 2017.

\bibitem{wang2014noncontact}
S.~Wang and K.~V. Larin, ``Noncontact depth-resolved micro-scale optical
  coherence elastography of the cornea,'' \emph{Biomedical Optics Express},
  vol.~5, no.~11, pp. 3807--3821, 2014.

\bibitem{manapuram2012vivo}
R.~K. Manapuram, S.~R. Aglyamov, F.~M. Monediado, M.~Mashiatulla, J.~Li, S.~Y.
  Emelianov, and K.~V. Larin, ``In vivo estimation of elastic wave parameters
  using phase-stabilized swept source optical coherence elastography,''
  \emph{Journal of biomedical optics}, vol.~17, no.~10, p. 100501, 2012.

\bibitem{de2018live}
V.~S. De~Stefano, M.~R. Ford, I.~Seven, and W.~J. Dupps~Jr, ``Live human
  assessment of depth-dependent corneal displacements with swept-source optical
  coherence elastography,'' \emph{PloS one}, vol.~13, no.~12, p. e0209480,
  2018.

\bibitem{zaitsev2013correlation}
V.~Y. Zaitsev, L.~Matveev, G.~Gelikonov, A.~Matveyev, and V.~Gelikonov, ``A
  correlation-stability approach to elasticity mapping in optical coherence
  tomography,'' \emph{Laser Physics Letters}, vol.~10, no.~6, p. 065601, 2013.

\bibitem{zaitsev2015}
V.~Y. Zaitsev, A.~L. Matveyev, L.~A. Matveev, G.~V. Gelikonov, V.~M. Gelikonov,
  and A.~Vitkin, ``Deformation-induced speckle-pattern evolution and
  feasibility of correlational speckle tracking in optical coherence
  elastography,'' \emph{Journal of biomedical optics}, vol.~20, no.~7, p.
  075006, 2015.

\bibitem{wang2007phase}
R.~K. Wang, S.~Kirkpatrick, and M.~Hinds, ``Phase-sensitive optical coherence
  elastography for mapping tissue microstrains in real time,'' \emph{Applied
  Physics Letters}, vol.~90, no.~16, p. 164105, 2007.

\bibitem{kennedy2012strain}
B.~F. Kennedy, S.~H. Koh, R.~A. McLaughlin, K.~M. Kennedy, P.~R. Munro, and
  D.~D. Sampson, ``Strain estimation in phase-sensitive optical coherence
  elastography,'' \emph{Biomedical optics express}, vol.~3, no.~8, pp.
  1865--1879, 2012.

\bibitem{chin2014analysis}
L.~Chin, A.~Curatolo, B.~F. Kennedy, B.~J. Doyle, P.~R. Munro, R.~A.
  McLaughlin, and D.~D. Sampson, ``Analysis of image formation in optical
  coherence elastography using a multiphysics approach,'' \emph{Biomedical
  optics express}, vol.~5, no.~9, pp. 2913--2930, 2014.

\bibitem{zaitsev2016optimized}
V.~Y. Zaitsev, A.~L. Matveyev, L.~A. Matveev, G.~V. Gelikonov, A.~A. Sovetsky,
  and A.~Vitkin, ``Optimized phase gradient measurements and phase-amplitude
  interplay in optical coherence elastography,'' \emph{Journal of biomedical
  optics}, vol.~21, no.~11, p. 116005, 2016.

\bibitem{matveyev2018vector}
A.~Matveyev, L.~Matveev, A.~Sovetsky, G.~Gelikonov, A.~Moiseev, and V.~Zaitsev,
  ``Vector method for strain estimation in phase-sensitive optical coherence
  elastography,'' \emph{Laser Physics Letters}, vol.~15, no.~6, p. 065603,
  2018.

\bibitem{zaitsev2016hybrid}
V.~Y. Zaitsev, A.~L. Matveyev, L.~A. Matveev, G.~V. Gelikonov, E.~V. Gubarkova,
  N.~D. Gladkova, and A.~Vitkin, ``Hybrid method of strain estimation in
  optical coherence elastography using combined sub-wavelength phase
  measurements and supra-pixel displacement tracking,'' \emph{Journal of
  biophotonics}, vol.~9, no.~5, pp. 499--509, 2016.

\bibitem{zaitsev2016robust}
V.~Y. Zaitsev, A.~L. Matveyev, L.~A. Matveev, G.~V. Gelikonov, E.~Gubarkova,
  N.~D. Gladkova, and A.~Vitkin, ``{Robust strain mapping in optical coherence
  elastography by combining local phase-resolved measurements and cumulative
  displacement tracking},'' in \emph{Optical Elastography and Tissue
  Biomechanics III}, vol. 9710.\hskip 1em plus 0.5em minus 0.4em\relax SPIE,
  2016, pp. 66 -- 74.

\bibitem{j2011recent}
T.~J~Hall, P.~E~Barboneg, A.~A~Oberai, J.~Jiang, J.-F. Dord, S.~Goenezen, and
  T.~G~Fisher, ``Recent results in nonlinear strain and modulus imaging,''
  \emph{Current medical imaging reviews}, vol.~7, no.~4, pp. 313--327, 2011.

\bibitem{Rivaz11_TMI}
H.~Rivaz, E.~M. Boctor, M.~A. Choti, and G.~D. Hager, ``Real-time regularized
  ultrasound elastography,'' \emph{IEEE Transactions on Medical Imaging},
  vol.~30, no.~4, pp. 928--945, 2011.

\bibitem{yuan2015analytical}
L.~Yuan and P.~C. Pedersen, ``Analytical phase-tracking-based strain estimation
  for ultrasound elasticity,'' \emph{IEEE transactions on ultrasonics,
  ferroelectrics, and frequency control}, vol.~62, no.~1, pp. 185--207, 2015.

\bibitem{ISBI2018}
H.~{Khodadadi}, A.~G. {Aghdam}, and H.~{Rivaz}, ``Direct strain estimation in
  ultrasound elastography using a novel dynamic programming approach,'' in
  \emph{2018 IEEE 15th International Symposium on Biomedical Imaging (ISBI
  2018)}, 2018, pp. 1182--1186.

\bibitem{zaitsev2014model}
V.~Y. Zaitsev, L.~Matveev, A.~Matveyev, G.~Gelikonov, and V.~Gelikonov, ``A
  model for simulating speckle-pattern evolution based on close to reality
  procedures used in spectral-domain oct,'' \emph{Laser Physics Letters},
  vol.~11, no.~10, p. 105601, 2014.

\bibitem{bashkatov2005optical}
A.~Bashkatov, E.~Genina, V.~Kochubey, and V.~Tuchin, ``Optical properties of
  human skin, subcutaneous and mucous tissues in the wavelength range from 400
  to 2000 nm,'' \emph{Journal of Physics D: Applied Physics}, vol.~38, no.~15,
  p. 2543, 2005.

\bibitem{goodman2005introduction}
J.~W. Goodman, \emph{Introduction to Fourier optics}.\hskip 1em plus 0.5em
  minus 0.4em\relax Roberts and Company Publishers, 2005.

\bibitem{bellman1954}
R.~Bellman, ``The theory of dynamic programming,'' \emph{Bulletin of the
  American Mathematical Society}, vol.~60, no.~6, pp. 503--515, 1954.

\bibitem{Kasai}
C.~{Kasai} and K.~{Namekawa}, ``Real-time two-dimensional blood flow imaging
  using an autocorrelation technique,'' in \emph{IEEE 1985 Ultrasonics
  Symposium}, 1985, pp. 953--958.

\end{thebibliography}

\end{document}